\documentclass[default]{aastex61}
\usepackage{graphicx,wrapfig}
\graphicspath{{./img/}}
\usepackage{amsmath}
\usepackage{amssymb}
\usepackage{ctable}
\usepackage{bm}

\newcommand{\sgra}{Sgr~A*}
\newcommand{\be}{\begin{equation}}
\newcommand{\ee}{\end{equation}}
\newcommand{\gcmag}{J1745$-$2900}
\newcommand{\xtemag}{XTE\,J1810$-$197}
\newcommand{\jmag}{J1622$-$4950}
\newcommand{\emag}{1E\,1547.0$-$5408}

\newcommand{\benum}{\begin{enumerate}}
\newcommand{\eenum}{\end{enumerate}}

\begin{document}

\title{VLA Observations of Single Pulses from the Galactic 
       Center Magnetar}
\shorttitle{Single Pulses from the GC Magnetar}
\shortauthors{Wharton et al.}

\author{R.~S.~Wharton}
\affiliation{Cornell Center for Astrophysics and Planetary Science 
             and Department of Astronomy, 
             Cornell University, Ithaca, NY 14853, USA}
\affiliation{Max-Planck-Institut f\"ur Radioastronomie, 
             Auf dem H\"ugel 69, D-53121 Bonn, Germany}

\author{S.~Chatterjee}
\affiliation{Cornell Center for Astrophysics and Planetary Science 
             and Department of Astronomy, 
             Cornell University, Ithaca, NY 14853, USA}

\author{J.~M.~Cordes}
\affiliation{Cornell Center for Astrophysics and Planetary Science 
             and Department of Astronomy, 
             Cornell University, Ithaca, NY 14853, USA}

\author{G.~C.~Bower}
\affiliation{Academia Sinica Institute of Astronomy and Astrophysics, 
             645 N. A'ohoku Place, Hilo, HI 96720, USA}

\author{B.~J.~Butler}
\affiliation{National Radio Astronomy Observatory, Socorro, NM 87801, USA}

\author{A.~T.~Deller}
\affiliation{Centre for Astrophysics and Supercomputing, 
             Swinburne University of Technology, 
             P.O. Box 218, Hawthorn, VIC 3122, Australia}

\author{P.~Demorest}
\affiliation{National Radio Astronomy Observatory, Socorro, NM 87801, USA}

\author{T.~J.~W.~Lazio}
\affiliation{Jet Propulsion Laboratory, California Institute of Technology,
             Pasadena, CA 91109, USA}

\author{S.~M.~Ransom}
\affiliation{National Radio Astronomy Observatory, 
             Charlottesville, VA 22903, USA}

\date{\centering \today}

\correspondingauthor{R.~S.~Wharton}
\email{wharton@mpifr-bonn.mpg.de}

\begin{abstract}
\noindent We present the results of a 7--12~GHz phased-array study 
of the Galactic center magnetar \gcmag\ with the 
Karl G. Jansky Very Large Array (VLA).  Using data from two 6.5~hour 
observations from September~2014, we find that the average profile is 
comprised of several distinct components at these epochs and is stable 
over $\sim$day timescales and $\sim$GHz frequencies. Comparison with 
additional phased VLA data at 8.7~GHz shows significant profile 
changes on longer timescales. 
The average profile at 7--12~GHz is dominated by the jitter of relatively 
narrow pulses. The pulses in each of the four main profile components seen 
in September~2014 are uncorrelated in phase and amplitude, though there 
is a small but significant correlation in the occurrence of pulses in two 
of the profile components. Using the brightest pulses, we 
measure the dispersion and scattering parameters of \gcmag. A joint fit 
of 38 pulses gives a 10~GHz pulse broadening time of 
$\tau_{\rm sc, 10} = 0.09 \pm 0.03~\rm ms$ and a dispersion measure of 
${\rm DM} = 1760^{+2.4}_{-1.3}~{\rm pc~cm}^{-3}$. Both of these results 
are consistent with previous measurements, which suggests that the 
scattering and dispersion measure of \gcmag\ may be stable on timescales 
of several years.  

\end{abstract}


\section{Introduction}
\label{sec:intro}

The Galactic center magnetar \gcmag\ is one of only four magnetars known 
to produce pulsed radio emission.  Like the other three radio-emitting 
magnetars, \xtemag, \emag, and \jmag\ \citep{crh06, crh07, lbb10}, \gcmag\ 
shows bright spiky emission with a flat spectral index and an 
integrated pulse profile that varies substantially on timescales of 
weeks to months \citep{lak15, tek15, tde17}. Careful study of these 
objects in their active radio state will reveal what relation they have 
to canonical radio pulsars.

Since \gcmag\ is only $\Delta \theta \approx 2\farcs4$ 
(projected distance of ${\sim}0.1~{\rm pc}$ at 8.5~kpc) from \sgra, 
it is also an excellent source to characterize the magneto-ionic 
environment along the line of sight to the Galactic center.  
Observations at radio frequencies have already found that \gcmag\ has 
the highest dispersion measure (DM) and rotation measure (RM) of any 
known pulsar \citep{sj13, efk13}.  Multi-frequency measurements of the 
pulse broadening time (caused by multipath scattering) have shown 
that the 1~GHz pulse broadening time is 
$\tau_{1\,{\rm GHz}} = 1.3\pm 0.2\,{\rm s}$ \citep{sle14}, which is 
almost three orders of magnitude less than previously expected 
\citep{lc98II}. By combining the time-domain scattering measurements of 
\citet{sle14} with VLBA imaging measurements of the angular broadening 
of \gcmag, \citet{bdd14} found that most of the scattering 
material is located far from the Galactic center.  Understanding the 
scattering along the line of sight to the Galactic center is essential 
for conducting searches for pulsars around \sgra.

To study the radio emission of \gcmag\ and measure the dispersion and 
scattering parameters along the line of sight to the Galactic center, 
we have conducted a single pulse analysis using data taken with the 
Karl G. Jansky Very Large Array (VLA) in a new phased-array pulsar mode. 
This new observing mode allows for large bandwidths 
(e.g., $\nu_{\rm obs} = 7-12~{\rm GHz}$), making the VLA the most 
sensitive radio telescope for Galactic center pulsar observations at 
these frequencies ($\nu \sim 10~\rm GHz$). 
The rest of the paper is organized as follows.  
In Section~\ref{sec:obs}, we discuss the observations. 
In Section~\ref{sec:prof_evol}, we explore the time and frequency 
evolution of the observed average profile and describe how it fits 
in the context of multi-epoch observations of \gcmag.  
In Section~\ref{sec:fluctuations}, we characterize the sub-pulses 
in each of the profile components of \gcmag\ and quantify the 
effects of rotational phase jitter. 
In Section~\ref{sec:dispersion}, we measure the dispersion and scattering 
parameters of \gcmag\ and in Section~\ref{sec:discussion} we discuss 
our results.

\section{Observations}
\label{sec:obs}

As part of a search for radio pulsars in the immediate vicinity 
of \sgra, we observed the Galactic center with the VLA in a new 
phased-array pulsar observing mode.  The phased-array pulsar 
mode uses the 
YUPPI\footnote{YUPPI (the ``Y'' Ultimate Pulsar Processing Instrument) 
is based on software developed for GUPPI \citep[the Green Bank Ultimate 
Processing Instrument,][]{drd08}.} 
software backend to produce either channelized time series data 
(for searching) or folded profiles (for pulsar timing).  YUPPI 
collects the coherently summed voltages from the VLA correlator 
and uses the DSPSR software package \citep{vsb11} to channelize 
or fold the data. The processing is done in real time on the 
correlator backend (CBE) computing cluster at the VLA.  YUPPI 
is a versatile pulsar instrument that allows for wide bandwidths 
(the full band for many receivers) and a variety of time and frequency 
resolution settings.  More details on the Galactic center search 
and the new pulsar processing mode will be provided in an upcoming 
paper (Wharton et al., in prep).

Phased-array observations were conducted during the transition from 
D$\rightarrow$DnC configuration on two consecutive days 
(2014~September~15$-$16, MJD~56915$-$6) for 6.5~hours per day. Each 
observation consisted of alternating scans of 600\,s on \sgra\ followed 
by 100\,s scans on the phase calibrator J1744$-$3116.  No polarization 
or flux density calibrators were observed. Data were recorded as summed 
polarizations using 4096~MHz of simultaneous bandwidth in two 2048~MHz 
windows centered on 8.2~GHz and 11.1~GHz to avoid very strong radio 
frequency interference (RFI) at 9.6~GHz.  The time and frequency 
resolution were set to 
$\delta t = 200~\mu \rm s$ and $\Delta \nu = 4~\rm MHz$ based on the 
considerations of a Galactic center pulsar search. Observational 
parameters are summarized in Table~\ref{tab:obs_param}.  

The phasing and processing of the phased-array data is done 
independently on small sub-bands, which are combined to produce 
the final data set.  For the 7--12~GHz search data, the 4096~MHz
band was processed in $32\times128$~MHz sub-bands. Phasing gain 
solutions are calculated independently for each sub-band during each 
phase calibration scan.  This can lead to amplitude offsets in both 
frequency and time.  To remove these offsets, we calculate a running 
10~second mean and standard deviation and rescale each channel to 
have zero mean and unit variance. 

In addition to the Galactic center search data, we also use phased VLA 
data obtained commensally during the Very Long Baseline Interferometry 
(VLBI) observations presented in \citet{bdd14, bdd15}. These 
observations were conducted at 8.7 GHz with a spanned bandwidth of 
256 MHz and typically lasted six hours.

\begin{table}[h]
\centering
\begin{tabular}{lr}
\toprule
\toprule
\multicolumn{2}{c}{Observational Parameters} \\
\midrule
Obs Date (MJD)                  &  56915.9~/~56916.9   \\
Time On Source ($T_{\rm obs}$)  & 5.2~hr~/~5.4~hr      \\
Sample Time ($\Delta t$)        & 200 $\mu {\rm s}$    \\
Frequency Coverage              & 7.1 -- 9.2, 10.0 -- 12.1~GHz \\
Frequency Channels              & $1024\times 4$~MHz   \\
Configuration                   & D$\rightarrow$DnC    \\
Beam Size ($\theta_{\rm HPBW}$) & 7\farcs2 \\
\bottomrule
\end{tabular}
\caption{\footnotesize Parameters for the phased-array VLA 
         observations conducted on 2014~Sep~15 and 2014~Sep~16.  
         Parameters that differ between the days are given as 
         Day 1~/~Day 2 in the rightmost column.}
\label{tab:obs_param}
\end{table}

\section{Profile Evolution}
\label{sec:prof_evol}

For most radio pulsars, the mean pulse profile is remarkably 
stable in time as a result of the stability of the magnetic field 
that guides the radio emission \citep{hmt75}.  
Profile changes are often caused by changes in the structure or 
orientation of the magnetic field.  For example, the steady separation 
of two components in the profile of the Crab pulsar (B0531+21) is 
explained by the gradual drift of the magnetic field axis towards the 
equator \citep{lgw13}.  Profile changes are also seen in binary pulsars 
like B1913+16 where geodetic precession gradually changes the direction 
of the magnetic field axis \citep{kramer98}.  We examine the evolution 
of the mean pulse profile of \gcmag\ in both time and frequency.  

\subsection{Time Evolution}
\label{ssec:time_evol}

To generate a mean profile for each of our two observations, we 
de-disperse and fold the data at the appropriate dispersion measure 
(DM) and period for each epoch.  
We use a dispersion measure of $\rm DM = 1760~pc~cm^{-3}$ for both 
epochs based on the single pulse measurements that will be discussed 
in Section~\ref{sec:dispersion}.
The de-dispersed time series are then folded over a range of trial 
periods using the Fast Folding Algorithm \citep[FFA,][]{staelin69}.  
Taking the best-fit parameters to be those that maximize the 
signal-to-noise ratio ($\rm S/N$) of the folded profile, we find 
barycentric spin periods of $P = 3.76453(3)~\rm s$ 
for $\rm MJD = 56915$ and $P = 3.76453(4)~\rm s$ 
for $\rm MJD = 56916$. 

\begin{figure}[hp]
\centering
  \includegraphics[width=0.9\textwidth]{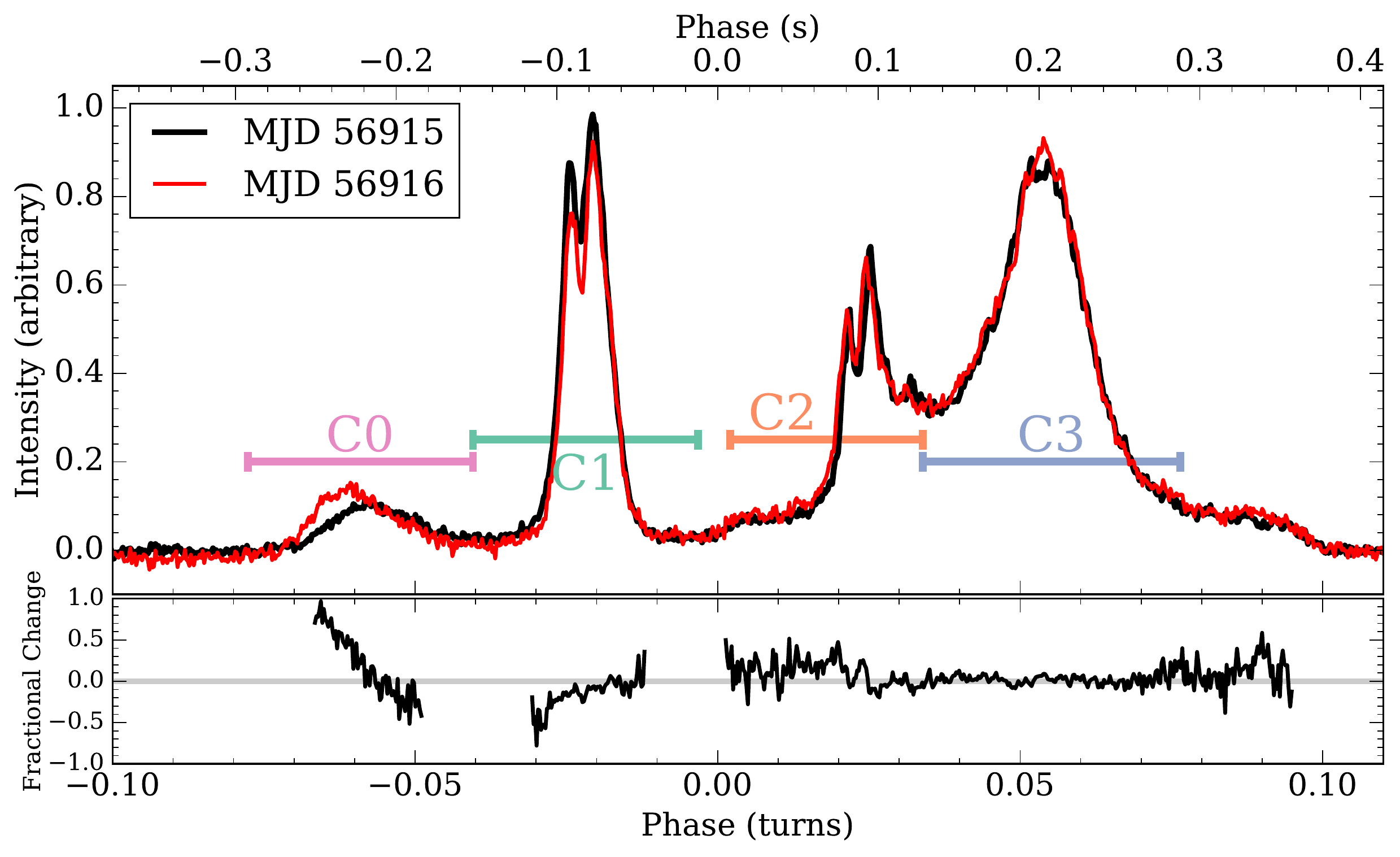}
\caption{\footnotesize \emph{Upper Panel:} Mean profiles of \gcmag\ 
observed on two consecutive days (MJD~56915, 56916).  The profiles 
have been normalized so that the area under each is the same.  The 
horizontal bars give the name and range of the four profile components.  
\emph{Lower Panel:}  The fractional difference between the two profiles.}
\label{fig:two_day}
\end{figure}

The mean profiles for each day are shown in Figure~\ref{fig:two_day}.  
They have been normalized so that the area under each pulse is the same, 
which allows for easier comparison.  We define four components 
(C0, C1, C2, C3) with widths of (140, 140, 120, 160)~ms that will 
be used throughout this paper.  
Though somewhat arbitrary, these components are useful for identifying 
the main regions from which single pulses arise.  
Both of the profiles show the same 
general structure with very similar substructure in each of the 
components. The main differences between the two are a slight amplitude 
change of C1 relative to C2~and~C3 and a shift in the peak of the 
relatively faint C0.  

\begin{figure}[hp]
\centering
  \includegraphics[width=0.45\textwidth]{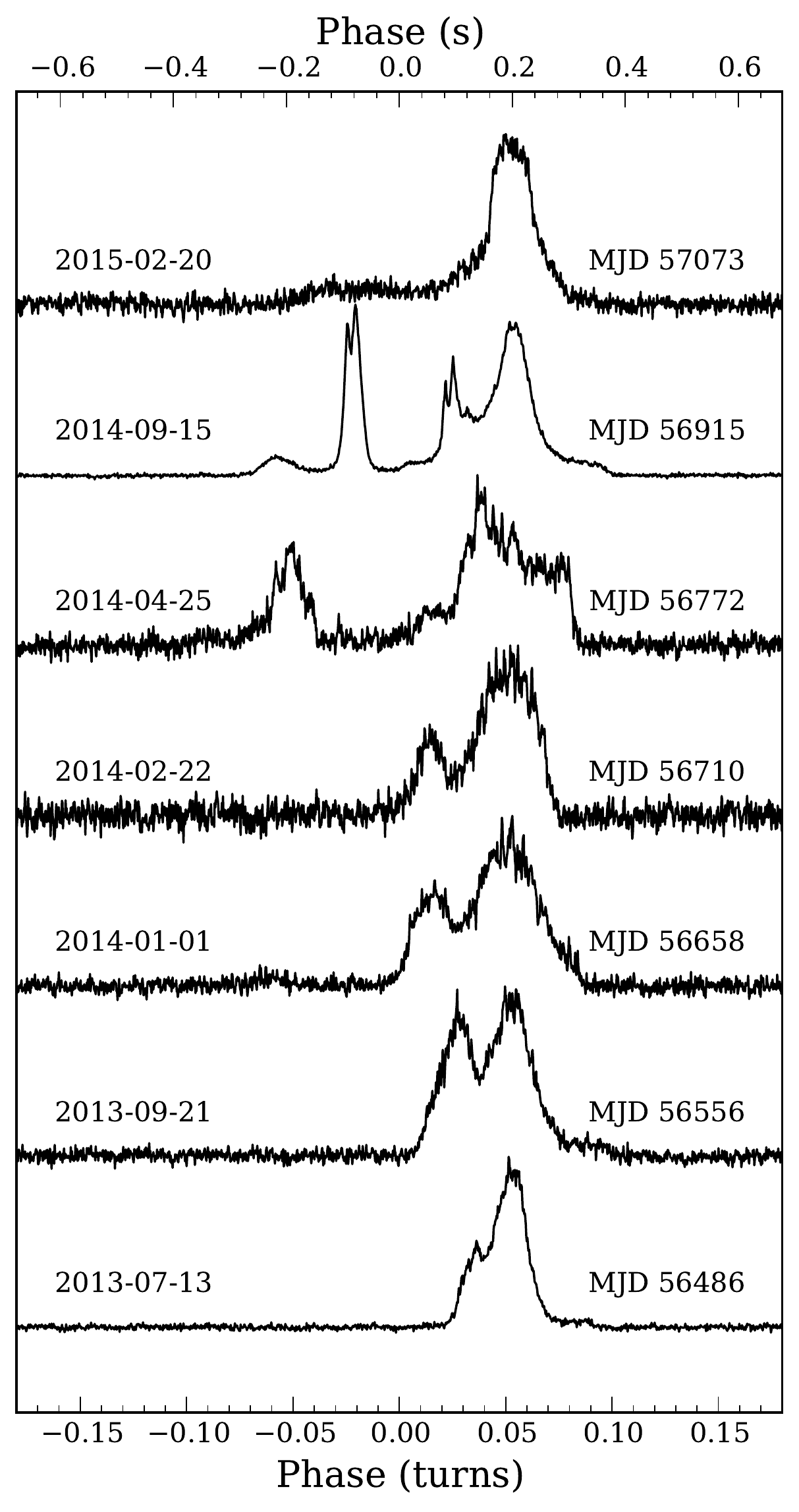}
\caption{\footnotesize 
Folded pulse profiles of \gcmag\ from phased-array VLA observations.  
The profile for MJD~56915 comes from a 6.5~hour observation using 
4~GHz of bandwidth.  The remaining profiles are from a VLBI campaign 
that used the phased VLA at 8.7~GHz using 256~MHz of bandwidth and 
observing times of about six hours \citep{bdd15}.  
The pulse profiles have been aligned so that the rightmost 
peak (our C3) of each pulse is roughly aligned. }
\label{fig:multi_epoch}
\end{figure}

While the mean profiles appear consistent over 1~day, this is not the 
case on longer time-scales.  Figure~\ref{fig:multi_epoch} shows a 
collection of \gcmag\ profiles generated from phased-array VLA data 
spanning ${\sim}600$~days. In addition to one of our profiles (MJD~56915), 
there are six profiles obtained commensally during VLBI observations of 
\gcmag\ using the phased VLA at 8.7~GHz with 256~MHz of bandwidth 
\citep{bdd14, bdd15}.  Since no phase-connected timing solution 
exists over this interval \citep{kab14, lak15}, we have simply aligned 
the profiles by the rightmost peak (our C3).

From Figure~\ref{fig:multi_epoch}, it is clear that \gcmag\ undergoes 
significant profile changes on long time scales. This behavior is 
consistent with the results from other monitoring campaigns.  
\citet{lak15} observed \gcmag\ with the GBT at 8.5~GHz once per week 
over the 167 days from MJD~56515--56682 and once per two weeks over 
the 130 days from MJD~56726--56856. 
During the earlier period (MJD~56515--56682), they found only minor 
changes to the mean profile as two components gradually separate. 
In the later period (MJD~56726--56856) the mean profile changes 
considerably, with components appearing and disappearing. Based on 
our VLA observations, it is likely that the period of gradual change 
extended at least until MJD~56710 (28 days beyond the last weekly GBT 
observation).  
Significant profile changes are also seen by \citet{ysw15} in six 
8.6~GHz observations with the Tian Ma Radio Telescope (TMRT) over 
the 107 day span from MJD~56836--56943.  Two of these observations 
occurred on consecutive days (MJD~56911, 56912) a few days before 
our observations.  These profiles are similar both to each other and 
to the profiles we observe on MJD~56915-6, although the much lower 
sensitivity prevents a more robust comparison. 
Profile stability on day time-scales is consistent with our 
observations (Figure~\ref{fig:two_day}).  Profile changes are 
also seen at frequencies of $89-291$~GHz \citep{tek15, tde17}, 
which strongly suggests a magnetospheric origin for these changes.

\subsection{Frequency Evolution}
\label{ssec:freq_evol}

\begin{figure}[hp]
\centering
  \includegraphics[width=0.9\textwidth]{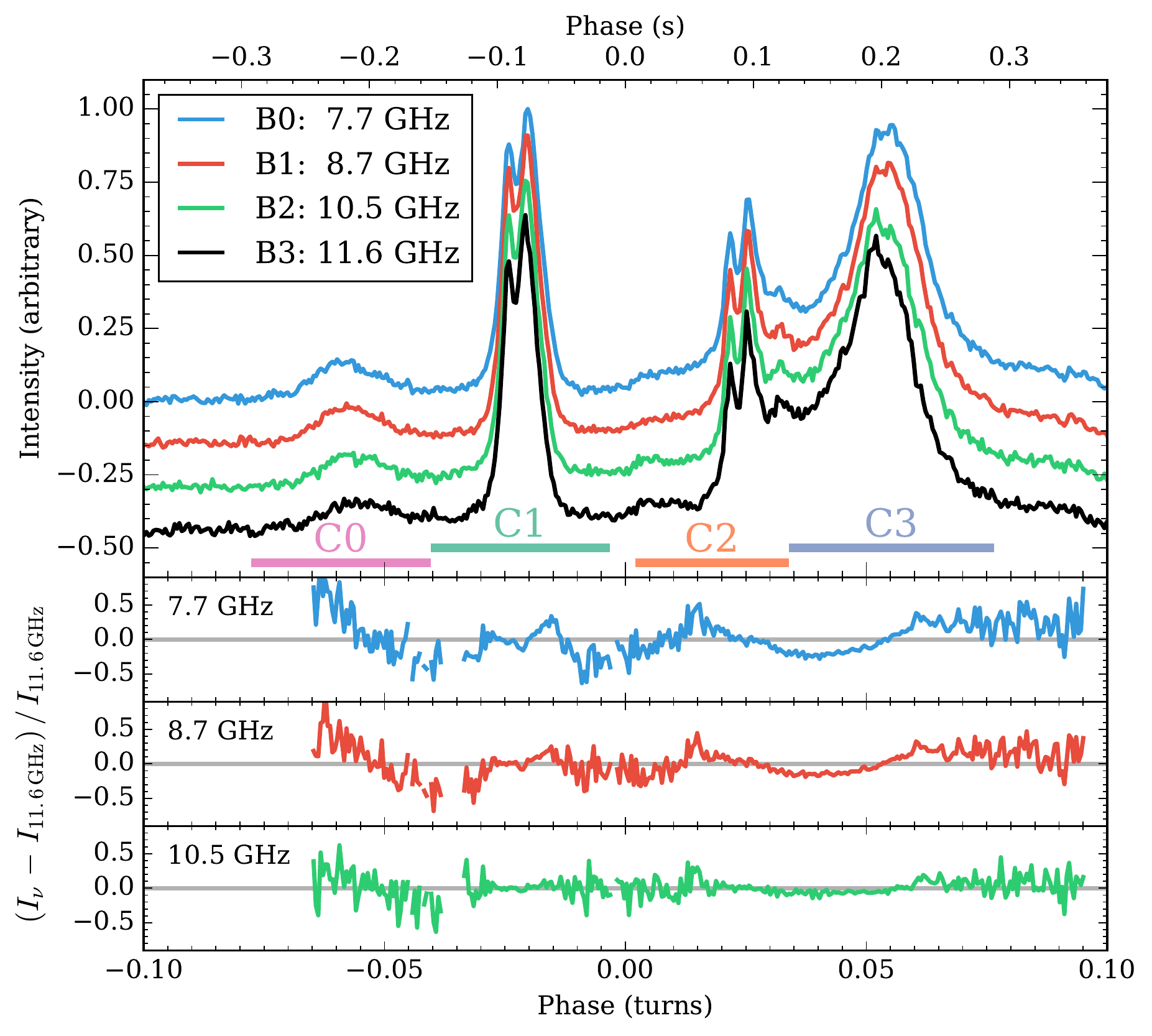}
\caption{\footnotesize 
Frequency evolution of \gcmag\ pulse profile for MJD~56915.  
\emph{Upper Panel:} Mean profiles generated using 1~GHz sub-bands 
(lowest frequency on top, highest on bottom). The center frequency 
of each sub-band is shown in the legend.  The profiles have been 
normalized so that the area under each pulse is the same.  Horizontal 
bars indicate the range of the profile components. 
\emph{Lower Panels:} Fractional difference between each of the 
three lower sub-bands and the sub-band centered on 11.6~GHz.}
\label{fig:subbands}
\end{figure}

Many pulsars show a gradual change in profile shape as a function 
of observing radio frequency \citep{thorsett91, cw14}.  
We can test whether there is a similar effect in our \gcmag\ data by 
splitting the 4~GHz bandwidth into four 1~GHz sub-bands (B0, B1, B2, B3) 
and generating mean profiles for each band. The resulting profiles for 
the MJD~56915 data set are shown in Figure~\ref{fig:subbands} along with 
the fractional difference between the profile generated from the highest 
frequency sub-band and the other three sub-bands.  

From Figure~\ref{fig:subbands}, we see that the mean profile of 
\gcmag\ is essentially consistent from 7.7--11.6~GHz, with a few 
slight changes.  For one, each of the peaks in the components 
C1, C2, and C3 narrow with increasing frequency. Another slight 
change is that the height of the bridge from C2 to C3 appears to be 
increasing with frequency, although this may be an artifact of the 
normalization of the pulses to equal area.  Finally, it seems as 
though the amplitude of C1 decreases with increasing frequency.  

The modest profile evolution in frequency seen here in \gcmag\ is 
consistent with that seen in radio pulsars at comparable frequencies 
\citep{kxj97, jkm08}.  In general, though, radio-emitting magnetars 
seem to show more complex behavior. \citet{kxj97} conducted a 
multifrequency study of \xtemag\ and found that the average profile 
could change significantly (e.g., appearance or disappearance of 
components) from $1.4-8.4~\rm GHz$.  Previous studies of \gcmag\ 
have also shown significant profile changes from $2.5-8.4~\rm GHz$ 
in some epochs \citep{tek15} and almost no frequency evolution 
in others \citep{tde17}.  This suggests that the frequency 
evolution of the average profile of \gcmag\ is time-dependent.

\vspace{1em}
\section{Single Pulse Properties}
\label{sec:fluctuations}

\begin{figure}[hp]
\centering 
  \includegraphics[width=1.0\textwidth]{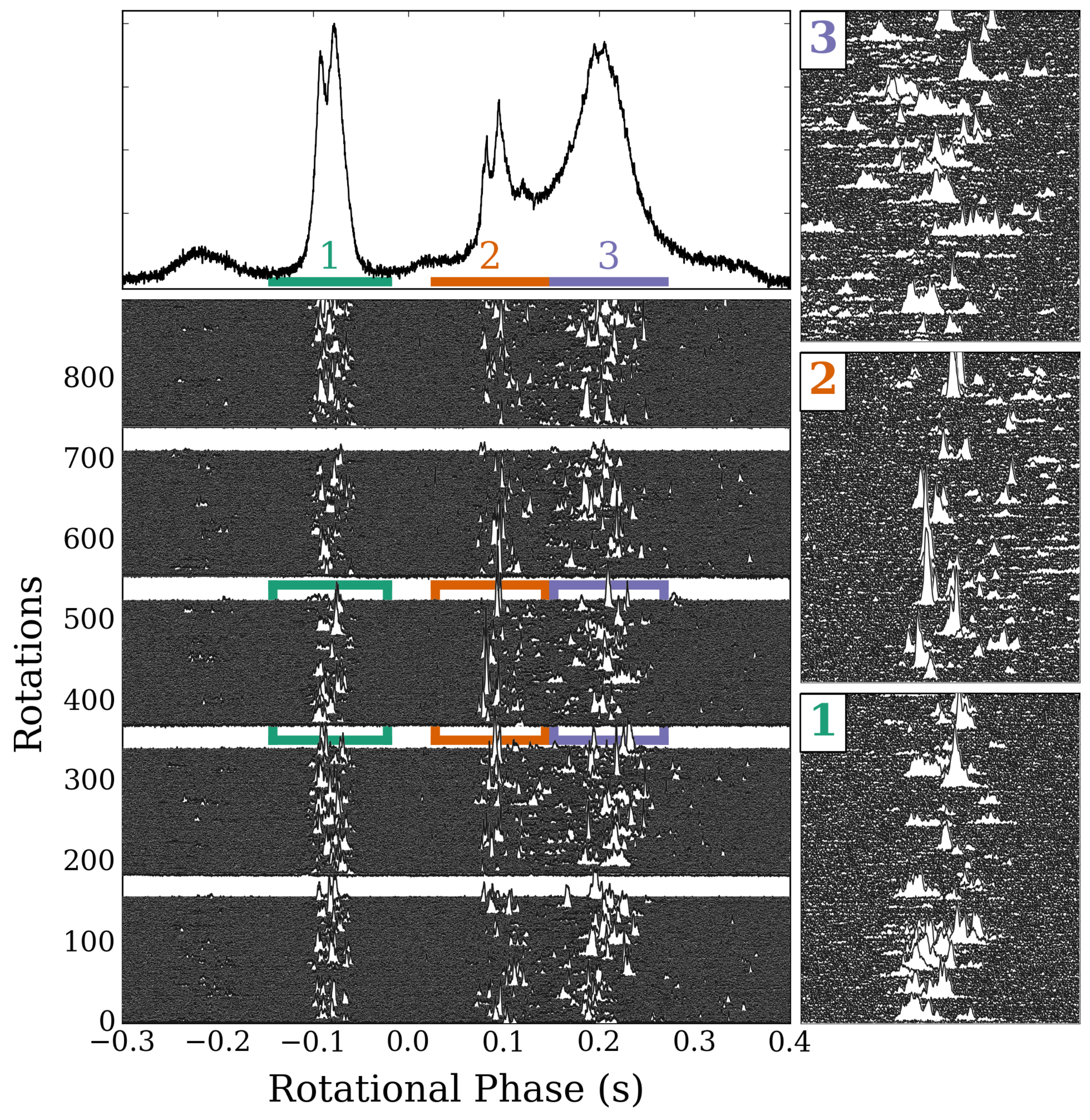}
\caption{\footnotesize 
Stacked single pulses from 900 rotations ($\approx 3400$~s) of \gcmag\ 
from MJD~56915. The 100~s ($\approx 27$~rotations) calibrator scans are 
seen as gaps in between the 600~s ($\approx 160$~rotations) on source 
scans. The upper panel shows the mean profile from the full observation. 
Panels on the right give a zoomed in view of the pulses showing 120~ms 
of rotational phase over one 600~s on source scan.}
\label{fig:stacked_sp}
\end{figure}

In each of our observations, we have collected single pulse data 
from nearly 5000 rotations of \gcmag.  Owing to the brightness 
of the magnetar and the sensitivity of the VLA, individual 
sub-pulses are clearly seen in almost every rotation.  
Figure~\ref{fig:stacked_sp} shows a selection of 900 rotations 
($\approx 3400$~s) of the magnetar from MJD~56915. 
The wide ($\approx 700$~ms) mean profile is comprised of much narrower  
($\sim 1-10$~ms) single pulses that appear to show a large degree of 
rotational phase jitter.  
As such, this is an excellent data set to quantify the jitter and 
to search for any correlations in the properties of sub-pulses 
occurring in each of the profile components.  Because the MJD~56916 
observation contained a significant amount of RFI, we only use single 
pulses from MJD~56915 in this analysis.

\subsection{Single Pulse Characterization}
\label{ssec:char}

To quantify the single pulse behavior of \gcmag,
we determine the amplitude, arrival time, and width of the 
pulse in each profile component for every rotation of the magnetar. 
We use a matched filter technique from pulsar timing in which the 
intensity, $I(t)$, of a pulse is represented as a scaled and shifted 
template, $G(t)$, in the presence of noise so that

\begin{equation}
    I(t) = b\, G(t-\tau) + c + n(t)
\end{equation}

where $b$ and $c$ are constants and $n(t)$ is noise.  The scale ($b$) 
and shift ($\tau$) parameters are found through fitting in the Fourier 
domain \citep{taylor92}.  We use a Python implementation of this fitting 
technique from the PyPulse software 
package\footnote{https://github.com/mtlam/PyPulse} \citep{lam17}. 

In pulsar timing, the template is typically taken to be the mean profile.  
The mean profile of \gcmag\ is far too wide to be useful for fitting the 
narrow pulses in each profile component, so we instead use Gaussians.  
Since the pulses appear to have a range of widths, we draw from a 
template bank of Gaussian functions with full-width at half-maximum (FWHM)
values of $w = 2^m \delta t$ for $m \in [1, 8]$ samples, which is 
$0.2-51.2~\rm ms$ for our time resolution of $\delta t = 0.2~\rm ms$.  
The width of the pulse corresponds to the width of the template that 
maximizes signal to noise ratio.  This is done for each profile component.
Thus, our fitting procedure returns an estimate for the amplitude 
($\hat{b}$ in units of the noise standard deviation), 
time-of-arrival offset ($\hat{\tau}$), and width ($\hat{w}$) of a pulse 
within each profile component for each rotation of the magnetar.

\subsection{Pulse Width Distribution}
\label{ssec:width}

\begin{figure}[hp]
\centering
  \includegraphics[width=0.9\textwidth]{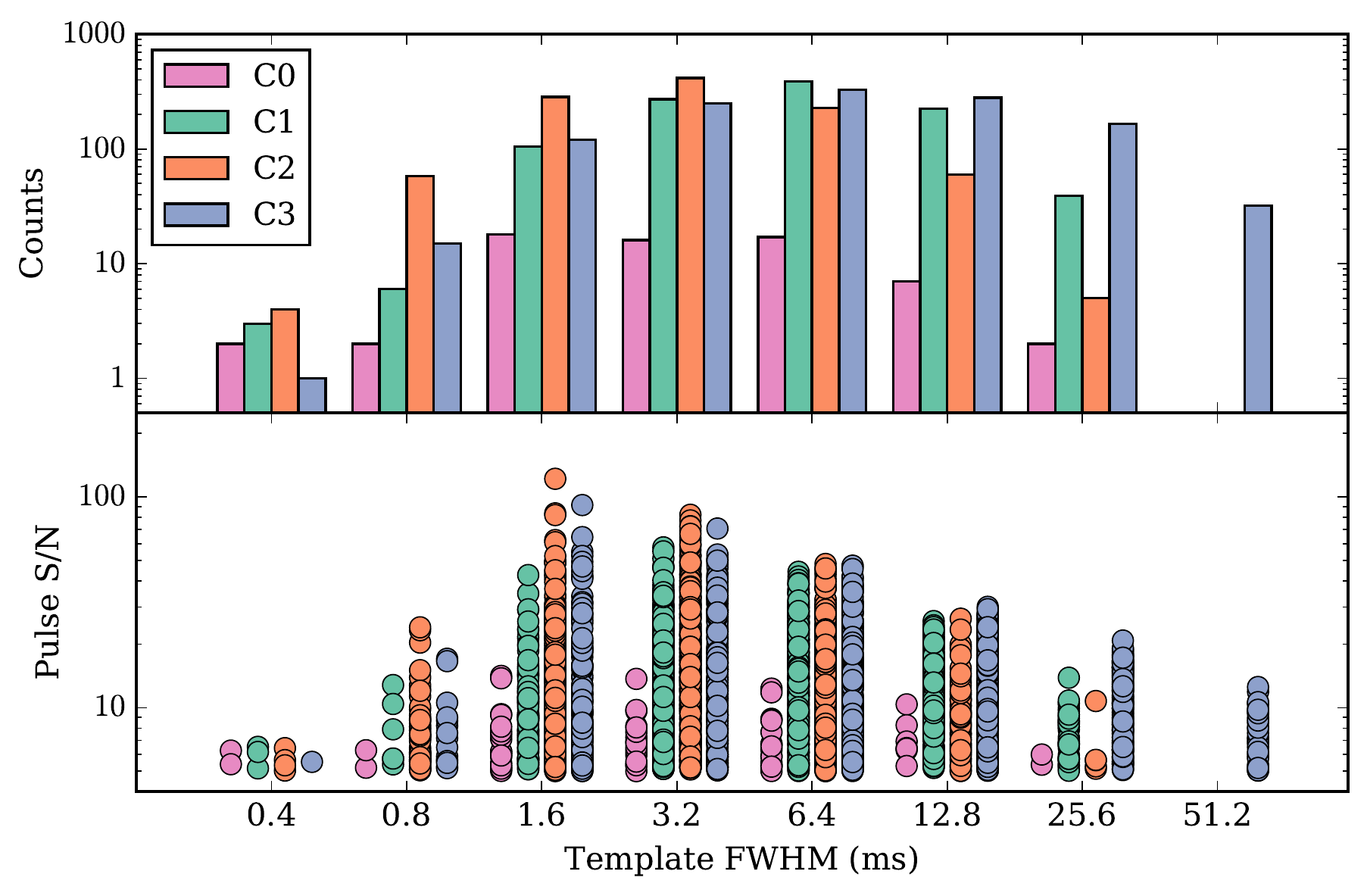}
\caption{\footnotesize Distribution of pulse widths for pulses  
         in each of the four profile components with a threshold 
         of ${\rm S/N} > 5$.
         \emph{Upper Panel:} Histogram of pulse widths for each 
         component. The total number of pulses in component 
         (C0, C1, C2, C3) is (64, 1038, 1053, 1191).  
         \emph{Lower Panel:} The $\rm S/N$ for each above-threshold 
         pulse as a function of width.   
        }
\label{fig:param_hist}
\end{figure}

It is clear from Figure~\ref{fig:stacked_sp} that individual pulses 
are seen with a variety of widths.  Using the measured widths from the 
template fitting, we determine the pulse width distribution for pulses 
from each profile component.  Figure~\ref{fig:param_hist} shows the width 
distribution for pulses with $\hat{b} > 5$ (that is, $\rm S/N > 5$). 
The most common pulse width for all components is either 3.2 or 6.4~ms.  
There are no pulses found with the narrowest template 
($\rm FWHM = 0.2~\rm ms$) and only in component C3 are there pulses found 
with the widest template ($\rm FWHM = 51.2~\rm ms$).  As seen in 
Figure~\ref{fig:stacked_sp}, these wide pulses are often comprised of 
many narrower (possibly overlapping) sub-pulses.

\subsection{Pulse Jitter}
\label{ssec:jitter}
Even though the mean profile of most pulsars is stable, the 
pulses from individual rotations can vary in both shape 
and arrival phase.  This phenomenon is called pulse jitter 
and is clearly present in the single pulses shown in 
Figure~\ref{fig:stacked_sp}. Following similar analyses in 
pulsar timing, we can estimate the contribution of pulse jitter 
to the overall time-of-arrival (TOA) measurement error.  Unlike 
most pulsar timing experiments, we will consider pulses from 
each profile component separately. The TOA measurement error, 
$\sigma_{\rm TOA}$, can be expressed as 

\begin{equation}
\sigma^2_{\rm TOA} = \sigma^2_{\rm S/N} + \sigma^2_{\rm DISS} + 
	                                 \sigma^2_{\rm J}
\end{equation}

where $\sigma_{\rm S/N}$ is the template fitting error, 
$\sigma_{\rm DISS}$ is the contribution to the uncertainty 
caused by diffractive interstellar scintillation (DISS), and 
$\sigma_{\rm J}$ is the pulse jitter \citep{cs10}.  By measuring 
or estimating $\sigma_{\rm TOA}$, $\sigma_{\rm S/N}$, 
and $\sigma_{\rm DISS}$, we can determine $\sigma_{\rm J}$.

The template fitting error, $\sigma_{\rm S/N}$, quantifies the 
contribution of purely additive noise to the timing error.  It 
depends on the pulse signal to noise ratio (S/N), so will vary 
from pulse to pulse, but for our data set we see typically see 
$\sigma_{\rm S/N} \lesssim 0.1~\rm ms$.  
Since we use a simple Gaussian template, there may also be some 
additional error caused by the slight differences between the 
template and the intrinsic pulse shape.  Based on the distribution 
of pulse widths (Figure~\ref{fig:param_hist}), we do not expect 
this to be more than $\sim 1~{\rm ms}$.

The DISS term is the result of averaging each pulse over a finite 
number of scintles in the time-frequency plane and can be estimated as 
$\sigma_{\rm DISS} \approx \tau_{\rm d} / \sqrt{N_{\rm s}}$, where 
$\tau_{\rm d}$ is the scattering time and $N_{\rm s}$ is the number 
of scintles.  The number of scintles is given by 

\begin{equation}
N_{\rm s} \approx 
      \left( 1 + \eta \frac{B}{\Delta \nu_{\rm d}}\right)
      \left( 1 + \eta \frac{T}{\Delta t_{\rm d}} \right)
\end{equation}

where $\eta$ is the scintle filling factor, $B$ is the bandwidth, 
$T$ is the integration time, $\Delta t_{\rm d}$ is the diffractive 
time-scale, and $\Delta \nu_{\rm d}$ is the diffractive bandwidth 
\citep{cl91}. 
The diffractive bandwidth is related to the scattering time as 
$\Delta \nu_{\rm d} = 1.16 / (2\pi \tau_{\rm d})$ \citep{cr98}.  
Using a 10~GHz scattering time of $\tau_{\rm d} = 0.1~\rm ms$ 
(Section~\ref{sec:dispersion}), we expect 
$\Delta \nu_{\rm d} \approx 1800~{\rm Hz}~(\tau_{\rm d}/0.1~\rm ms )^{-1}$.
The diffractive time-scale for a single thin scattering screen is 
$\Delta t_{\rm d} = \ell_{\rm d} / v 
                  = \lambda / (2\pi \theta_{\rm d} v)
                  \sim 1~\rm s$,
where $\lambda = 3.5~\rm cm$, $\theta_{\rm d} = 15~\rm mas$, and 
$v \sim 100~\rm{km~s}^{-1}$ \citep{bdd15}.  Taking $\eta \approx 0.3$, 
$T = 3.76~\rm s$, and $B = 4~\rm GHz$, we find that the DISS contribution 
to the TOA uncertainty is only $\sigma_{\rm DISS} \approx 0.1~\mu\rm s$.

The single pulse TOA measurement error, $\sigma_{\rm TOA}$, is simply 
the observed scatter in TOA offsets ($\hat{\tau}$) measured for the 
pulses in each profile component.  It varies between the components, 
but typical values are $\sigma_{\rm TOA} \gtrsim 20~\rm ms$.
Since $\sigma_{\rm TOA} \gg \sigma_{\rm S/N} \gg \sigma_{\rm DISS}$,  
the single pulse TOA measurement error in every component is entirely 
dominated by the jitter so $\sigma_{\rm J}\approx\sigma_{\rm TOA}$.  

\begin{figure}[hp]
\centering
  \includegraphics[width=0.9\textwidth]{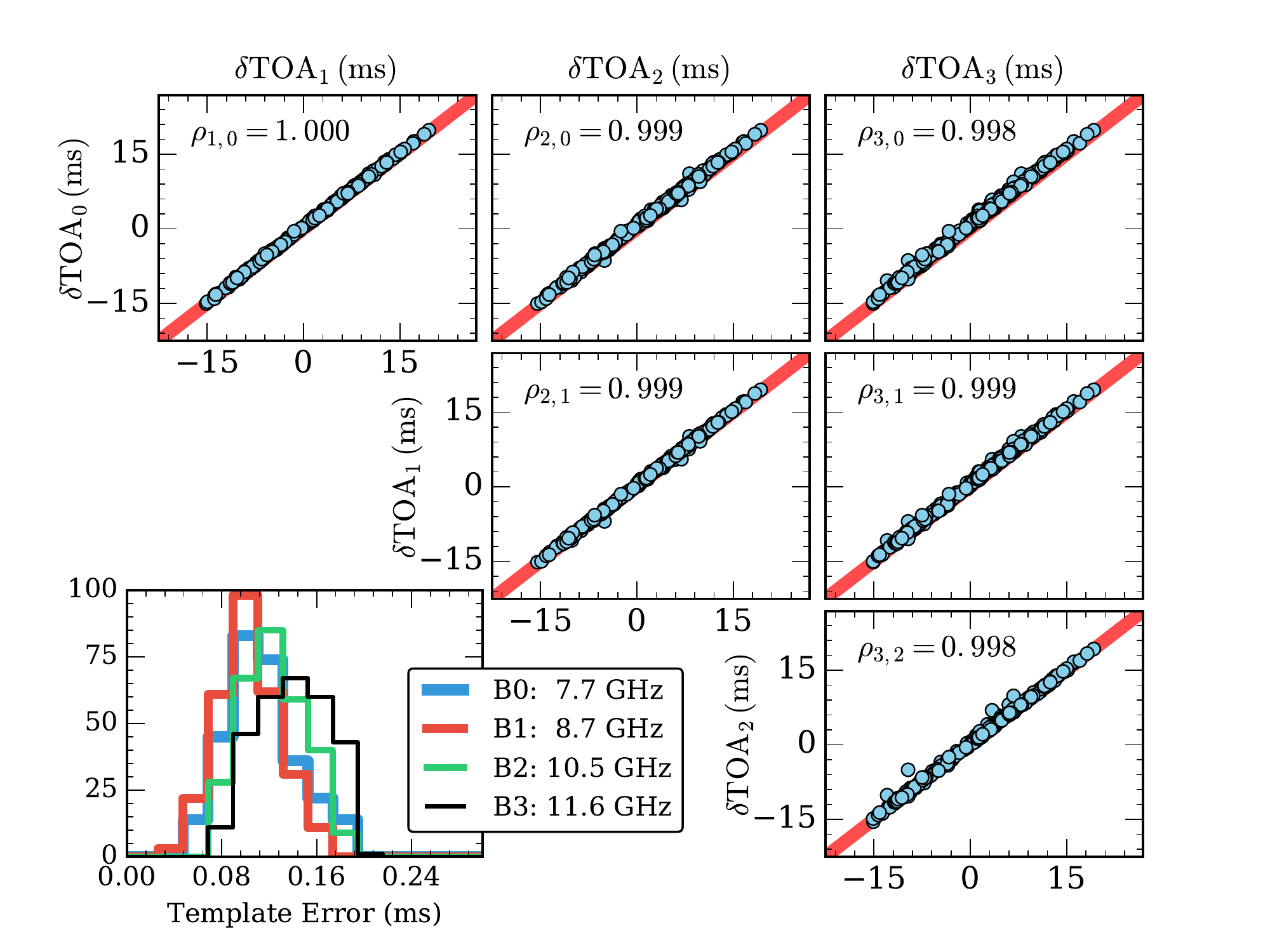}
\caption{\footnotesize Comparison of time-of-arrival (TOA) offsets 
for single pulses from profile component C1 measured in four 1~GHz 
sub-bands. We selected the 288 pulses detected with $\rm S/N > 5$ in 
all four bands and calculated the TOA offsets using a Gaussian template 
with a full-width at half-max of 10~ms. The correlation coefficient for 
each pairwise comparison of sub-bands is shown in each panel.  The lower 
left panel shows the distribution of template fitting errors, 
$\sigma_{\rm S/N}$, for all the pulses.}
\label{fig:toa_corr_mat}
\end{figure}

The pulse phase jitter is correlated over typical observing bands, but 
can decorrelate over larger bandwidths.  In a study of millisecond 
pulsars, \citet{sod14} found that the jitter in PSR~J0437$-$4715 
decorrelates between 0.75$-$3.1~GHz, giving a jitter correlation 
bandwidth of $\lesssim\!2$~GHz.  To test the jitter correlation bandwidth 
of \gcmag, we split the full 4~GHz band into four 1~GHz sub-bands 
(centered on frequencies of 7.7, 8.7, 10.5, and 11.6~GHz), characterize 
the pulses in each component for each sub-band, and then compare the 
results. Instead of searching over a range of pulse widths, we just use 
a single template with $\rm FWHM = 10~ms$ (50 time samples). This ensures 
a consistent comparison of pulses in different frequency bands.  
Figure~\ref{fig:toa_corr_mat} shows the TOA offsets measured in component 
C1 for all the sub-bands plotted against each other.  Plots from other 
profile components are similar. The pulse phase jitter in \gcmag\ is 
highly correlated over 4~GHz of bandwidth.

\subsection{Correlations between Profile Components}
\label{ssec:corr}

In Section~\ref{ssec:jitter}, we showed that the TOA measurement 
error of pulses within each profile component is dominated by jitter.  
Here we explore whether there are any correlations between the 
TOA offsets or amplitudes of these pulses. Any correlation in the 
properties of pulses between components could indicate a common 
origin for pulsed emission.  

To look for correlations, we first generate time series data for the 
pulse properties of interest from each profile component using the 
methods described in Section~\ref{ssec:char} and a threshold of 
$\rm (S/N)_{\rm min} = 5$. We determine the TOA offset ($\hat{\tau}[n]$), 
the pulse amplitude ($\hat{b}[n]$), and a binary value ($\Theta[n]$) 
indicating the presence of an above threshold pulse all as a function of 
the pulse number ($n$). Next, the cross-correlation function (CCF) is 
calculated between the time series of two profile components for each 
pulse parameter of interest.  We denote the CCF as 

\begin{equation}
{\rm CCF}_{ij}(x)[n] = \left( x_i \star x_j \right)[n]
\end{equation}

where $x$ is the time series parameter ($\hat{b}, \hat{\tau}, \Theta$)  
from the profile components $i,j \in \{\rm C0, C1, C2, C3\}$ normalized 
to have zero mean and unit variance.  To avoid the periodicity introduced 
by the calibrator scans, the CCFs are calculated using data from each 
on-source scan and then averaged together over all scans.  

To determine the significance of any CCF peaks, we shuffle all the values 
in the time series of each parameter and re-calculate the CCF, repeating 
this process $10^4$ times.  Since the shuffled time series should have no 
correlations, we can use these results to set the 99.7\% 
($\approx 3\sigma$) confidence level for any lag value in the CCFs.  

\begin{figure}[hp]
\centering
  \includegraphics[width=0.95\textwidth]{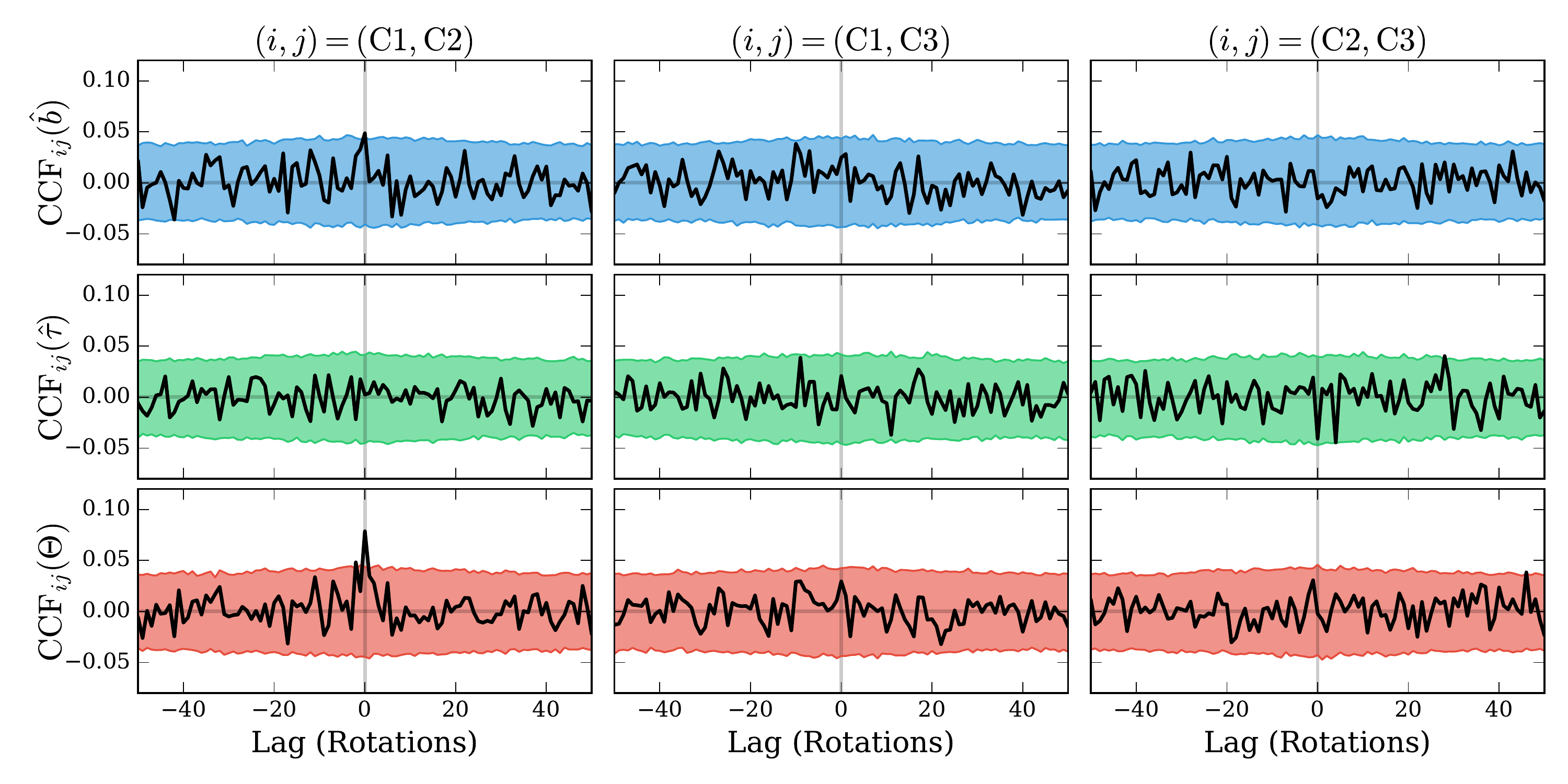}
\caption{\footnotesize 
Cross-correlation function (CCF) between the time series of 
pulse properties measured in the three brightest pulse profile 
components. The black lines denote the observed CCF and the light 
shaded regions denote the 99.7\% ($\approx 3\sigma$) confidence 
limit estimated by reshuffling the pulse order (see text for details).  
\emph{Upper Row:}  CCF of observed pulse amplitudes ($\hat{b}$).
\emph{Middle Row:} CCF of observed pulse TOA offsets ($\hat{\tau}$).
\emph{Lower Row:} CCF of above threshold pulse occurrence ($\Theta$).}
\label{fig:comp_ccfs}
\end{figure}

The results of this analysis are shown in Figure~\ref{fig:comp_ccfs}.  
We calculate ${\rm CCF}_{ij}(x)[n]$ for each parameter 
($\hat{b}, \hat{\tau}, \Theta$) for component pairs 
$(i, j) = \rm (C1,C2)$, $\rm (C1, C3)$, and $\rm (C2,C3)$.
Component C0 was excluded because it had far fewer above threshold 
pulses ($N=64$) than C1 ($N=1038$), C2 ($N=1053$), and C3 ($N=1191$).
While most of the CCFs appear to be consistent with noise, there is a 
small but significant correlation in the occurrence of pulses in C1 and 
C2 at zero lag. This means that pulses in C1 and C2 occur during the 
same rotation of \gcmag\ more often than would be expected if they were 
completely independent.

\vspace{1em}
\section{Dispersion and Scattering in Single Pulses}
\label{sec:dispersion}
As a bright radio-emitting magnetar in the immediate vicinity 
of \sgra, \gcmag\ is an excellent tool for studying the 
magneto-ionic environment along the line of sight to the 
Galactic center. Measurements of the dispersion measure and 
scattering time are easiest for bright and narrow pulse profiles. 
The broad jitter-dominated average profile of \gcmag\ is not well 
suited for these measurements at 10~GHz, but some of the individual 
sub-pulses are. Here we use a set of bright narrow sub-pulses 
(hereafter just referred to as pulses) to measure the dispersion 
and scattering parameters for \gcmag.

\subsection{Pulse Selection}
\label{ssec:pulses}
\begin{figure}
\centering

\begin{tabular}[b]{@{}p{0.42\textwidth}@{}}
\centering\includegraphics[width=\linewidth]{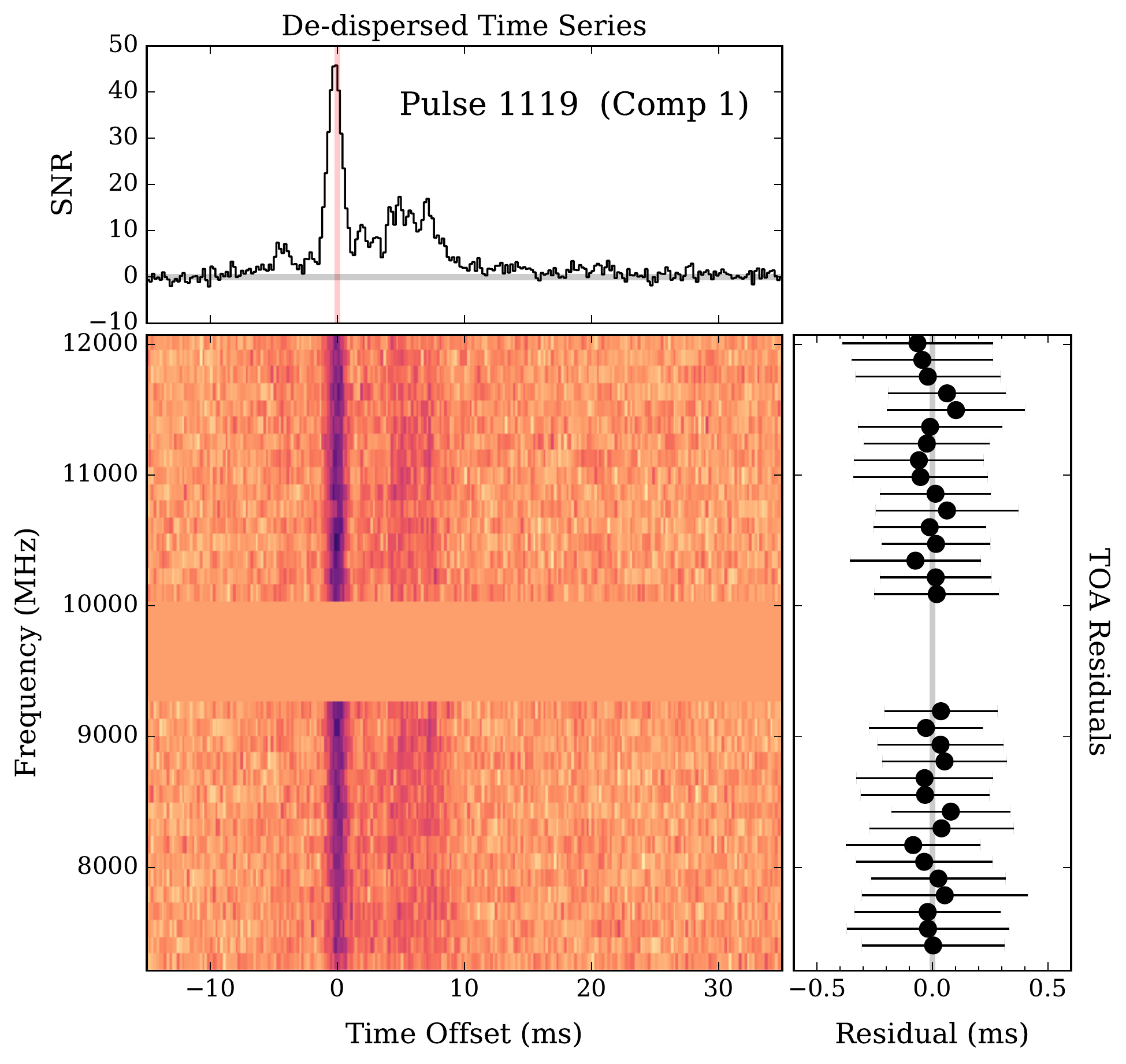} \\
\end{tabular}%
\begin{tabular}[b]{@{}p{0.42\textwidth}@{}}
\centering\includegraphics[width=\linewidth]{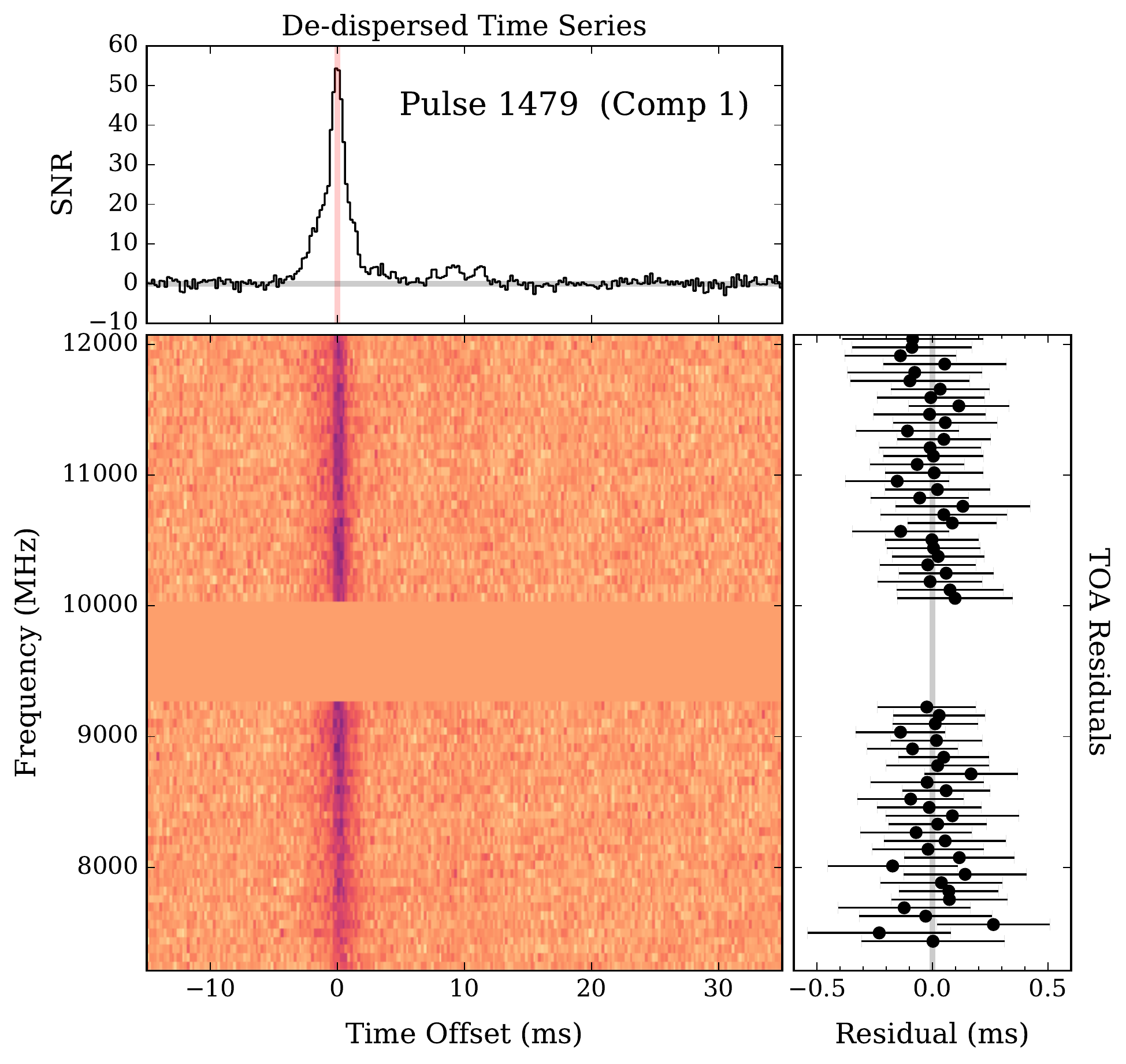} \\
\end{tabular}%

\vspace{-1em}
\begin{tabular}[b]{@{}p{0.42\textwidth}@{}}
\centering\includegraphics[width=\linewidth]{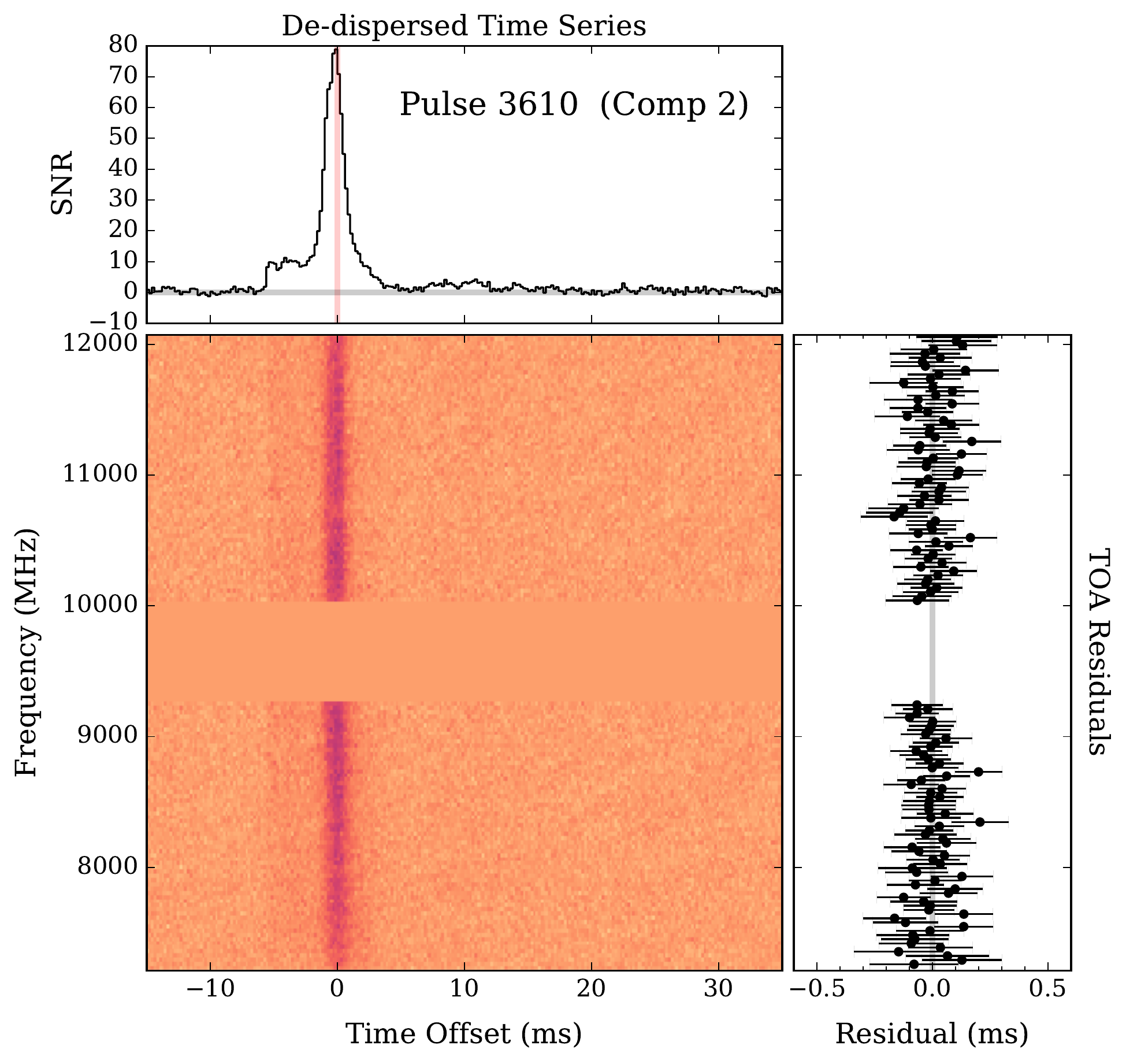} \\
\end{tabular}%
\begin{tabular}[b]{@{}p{0.42\textwidth}@{}}
\centering\includegraphics[width=\linewidth]{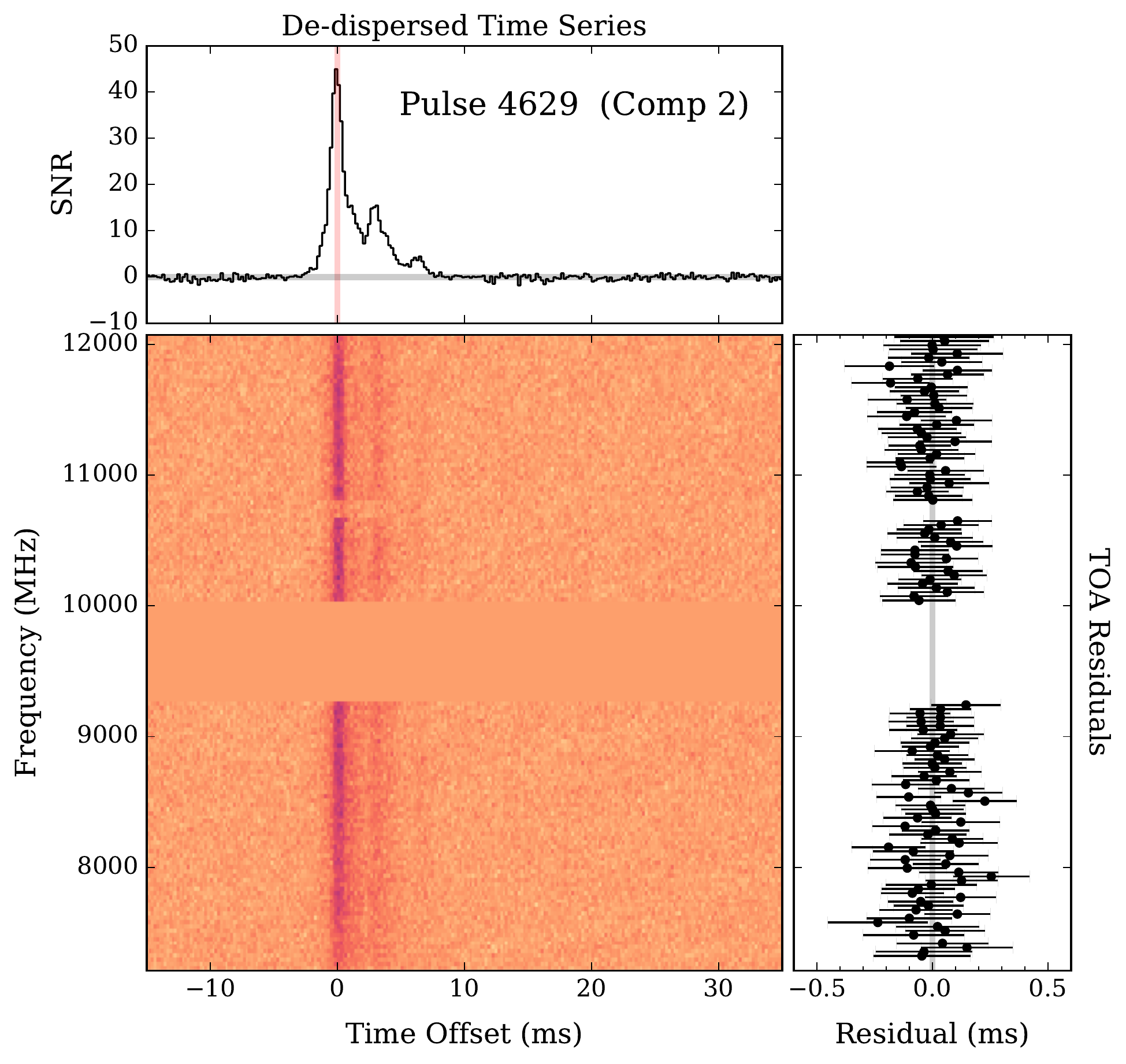} \\
\end{tabular}

\vspace{-1em}
\begin{tabular}[b]{@{}p{0.42\textwidth}@{}}
\centering\includegraphics[width=\linewidth]{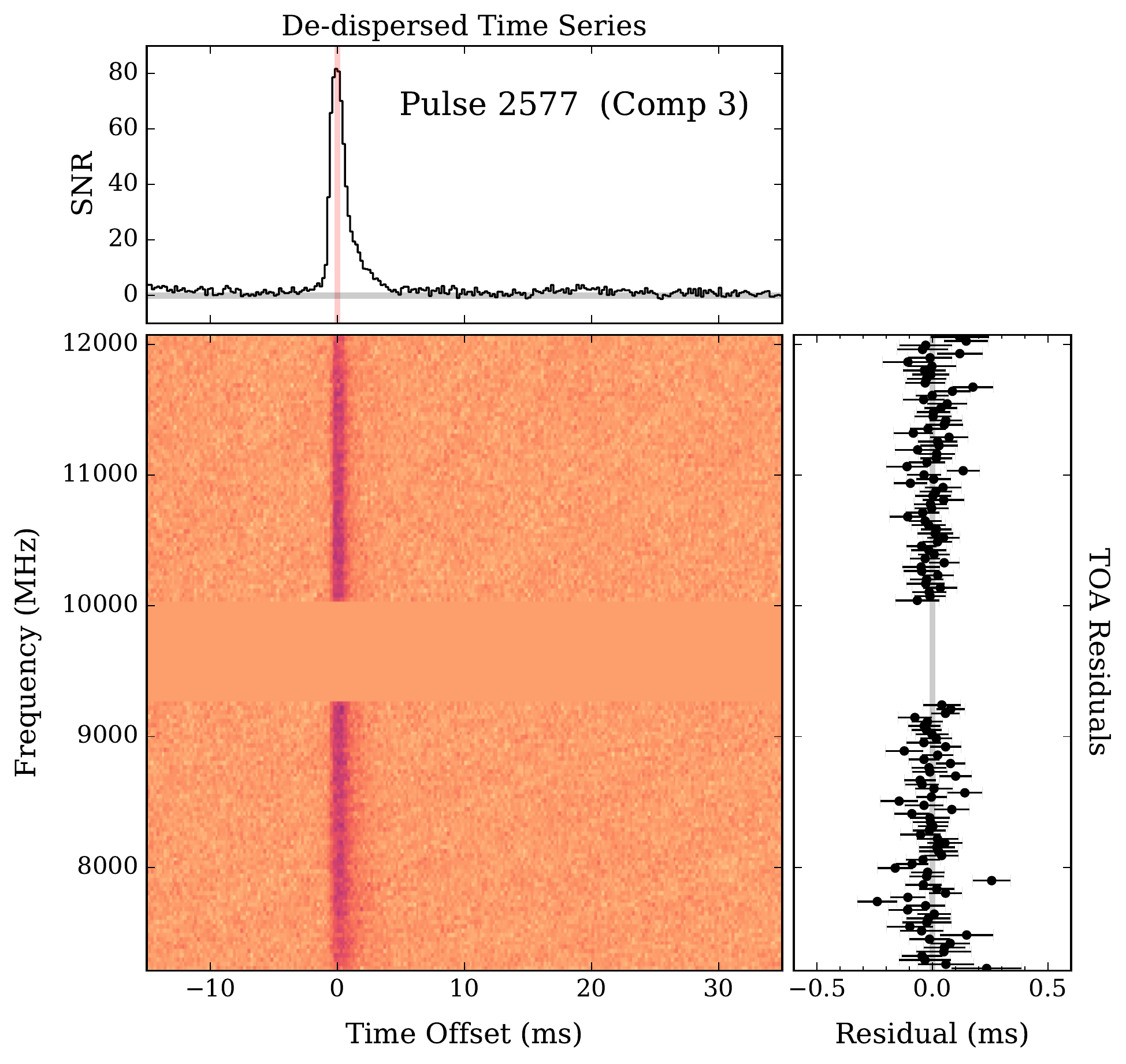} \\
\end{tabular}%
\begin{tabular}[b]{@{}p{0.42\textwidth}@{}}
\centering\includegraphics[width=\linewidth]{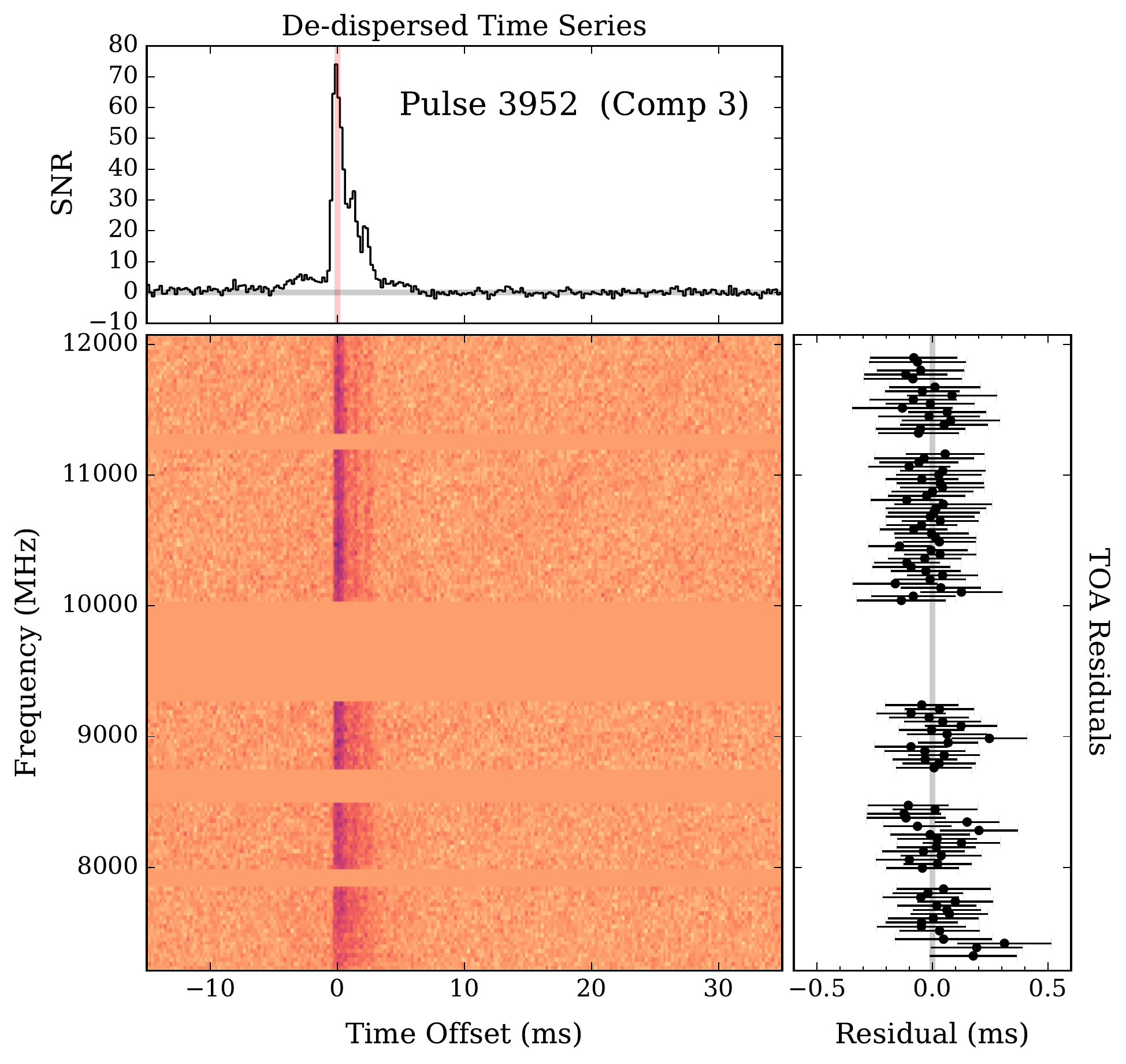} \\
\end{tabular}

\caption{A sample of six of the 38 pulses used to fit for the dispersion 
         measure and scattering. Each panel shows the dispersion corrected 
         frequency-resolved pulse (middle), the de-dispersed time series 
         (top), and residual delays after subtracting the global best-fit 
         model (right). The frequency range from 9.2--10~GHz was avoided 
         due to strong RFI and other gaps are the result of RFI masking.}
\label{fig:dspec_plots}
\end{figure}

To get the best measurements of the dispersion measure and 
pulse broadening time, we need to select pulses that have 
high S/N and small widths. Using the pulse parameters determined 
in Section~\ref{ssec:char}, we select pulses with $\rm S/N > 30$ 
and $\hat{w} \leq \rm 8~bins = 1.6~\rm ms$.  A total of 40 pulses 
meet these criteria, but two are excluded (one for missing data 
and one for having a wide and complicated pulse structure).  Of 
the 38 remaining pulses, none are from component C0, 2 are from C1, 
23 are from C2, and 13 are from C3.  Only one of the selected pulses 
has a width of $\hat{w} = 4~\rm bins$ (0.8~ms), the rest have a width 
of $\hat{w} = 8~\rm bins$ (1.6~ms).  The widths are only approximate, 
though, as they were found by matched filtering using Gaussian 
templates.  The actual pulses can show more complicated structure 
than the templates. Figure~\ref{fig:dspec_plots} shows the 
frequency-resolved and frequency-summed profiles for six of the 
38 pulses.

\subsection{Method}
\label{ssec:method}
Using the sample of 38 bright and narrow pulses, we can measure the 
dispersion and scattering by modeling the frequency-dependent delay 
across the observing band.  Dispersion introduces a delay in the 
arrival time of a pulse that scales as $\tau_{\rm DM} \propto \nu^{-2}$. 
Multipath scattering distorts the pulse so that the observed pulse 
shape is the intrinsic pulse shape convolved with a one-sided 
exponential with time-scale 
$\tau_{\rm sc} \propto \nu^{\alpha_{\rm sc}}$. This asymmetric 
distortion produces a frequency-dependent shift in the observed 
arrival time of a pulse. When the scattering time is small 
compared to the pulse width, the frequency-dependent shift in the 
measured arrival time of a pulse can be approximated as 

\begin{equation}
\label{eqn:toa_model}
   \tau(\nu) = k_{\rm DM} {\rm DM} \, 
      \left(\frac{\nu}{1~{\rm GHz}}\right)^{-2} 
    + \tau_{\rm sc, 10} \left(\frac{\nu}{10~{\rm GHz}}\right)^{\alpha_{\rm sc}} 
    + t_0
\end{equation}

where $k_{\rm DM} = 4.15~\rm ms~GHz^2~pc^{-1}~cm^3$ is the dispersive 
constant, ${\rm DM}$ is the dispersion measure, $\tau_{\rm sc, 10}$ is 
the scattering time at $10~\rm GHz$, $\alpha_{\rm sc}$ is the scaling 
index of the scattering law, and $t_0$ is an offset. 
The scattering index is fixed at $\alpha_{\rm sc} = -4.0$, 
which is consistent with the $\alpha = -3.8\pm 0.2$ measured for \gcmag\ 
by \citet{sle14}. We have found that the approximation of 
Equation~\ref{eqn:toa_model} is accurate to about 10\% for scattering 
times less than about 20\% the width of a pulse.  

Using the channelized time series data, sub-band arrival times 
for each pulse can be determined using the same matched filter 
method used in Section~\ref{ssec:char}.  For each pulse, we average 
together several frequency channels to ensure that $\rm S/N > 5$ 
detections can be made in each sub-band. Of the 38 pulses in our sample, 
(10, 16, 11, 1) pulses use sub-band bandwidths of (32, 64, 128, 256)~MHz.  
Using the measured arrival times and arrival time uncertainties in each 
sub-band, we fit for the parameters of the delay model 
(Equation~\ref{eqn:toa_model}) for each pulse separately and for all 
pulses jointly.  

Assuming normally distributed errors in the measured sub-band arrival 
times, the likelihood function for a single pulse fit is   

\begin{equation}
\mathcal{L}(\tau | {\rm DM}, \tau_{\rm sc,10}, t_0) = 
              \prod_{i=0}^{N-1} \frac{1}{\sqrt{2 \pi \sigma_{\tau, i}^2}}
              \exp\left[-\frac{1}{2} 
              \left(\frac{\tau_i - \tau(\nu_i)}{\sigma_{\tau,i}}\right)^2
              \right]
\end{equation}

where $\tau_i$ and $\sigma_{\tau,i}$ are the arrival time and 
arrival time uncertainty in the sub-band with center frequency 
$\nu_i$. Wide priors are adopted for each of the parameters.  
Normal distributions are used for the priors of 
$\rm DM$ ($\mu_{\rm DM} = 2000~{\rm pc~cm}^{-3}$, 
$\sigma_{\rm DM} = 1000~{\rm pc~cm}^{-3}$) 
and $t_0$ ($\mu_{t} = 0~{\rm ms}$, $\sigma_{t} = 500~{\rm ms}$).  
For the 10~GHz scattering time $\tau_{\rm sc, 10}$, an 
exponential distribution with mean $\lambda_{\tau} = 2~\rm ms$ 
is used as a prior.  Combining these priors with the likelihood, 
we can construct and sample the posterior distribution using the 
\texttt{emcee} MCMC sampler \citep{fhl13}. 
In addition to fitting each pulse separately for 
($\rm DM$, $\tau_{\rm sc,10}$, $t_0$), we can also do a global fit 
that assumes one set of ($\rm DM$, $\tau_{\rm sc,10}$) for all pulses. 

\subsection{Results}
\label{ssec:results}
\begin{figure}[hp]
\centering 
  \includegraphics[width=0.7\textwidth]{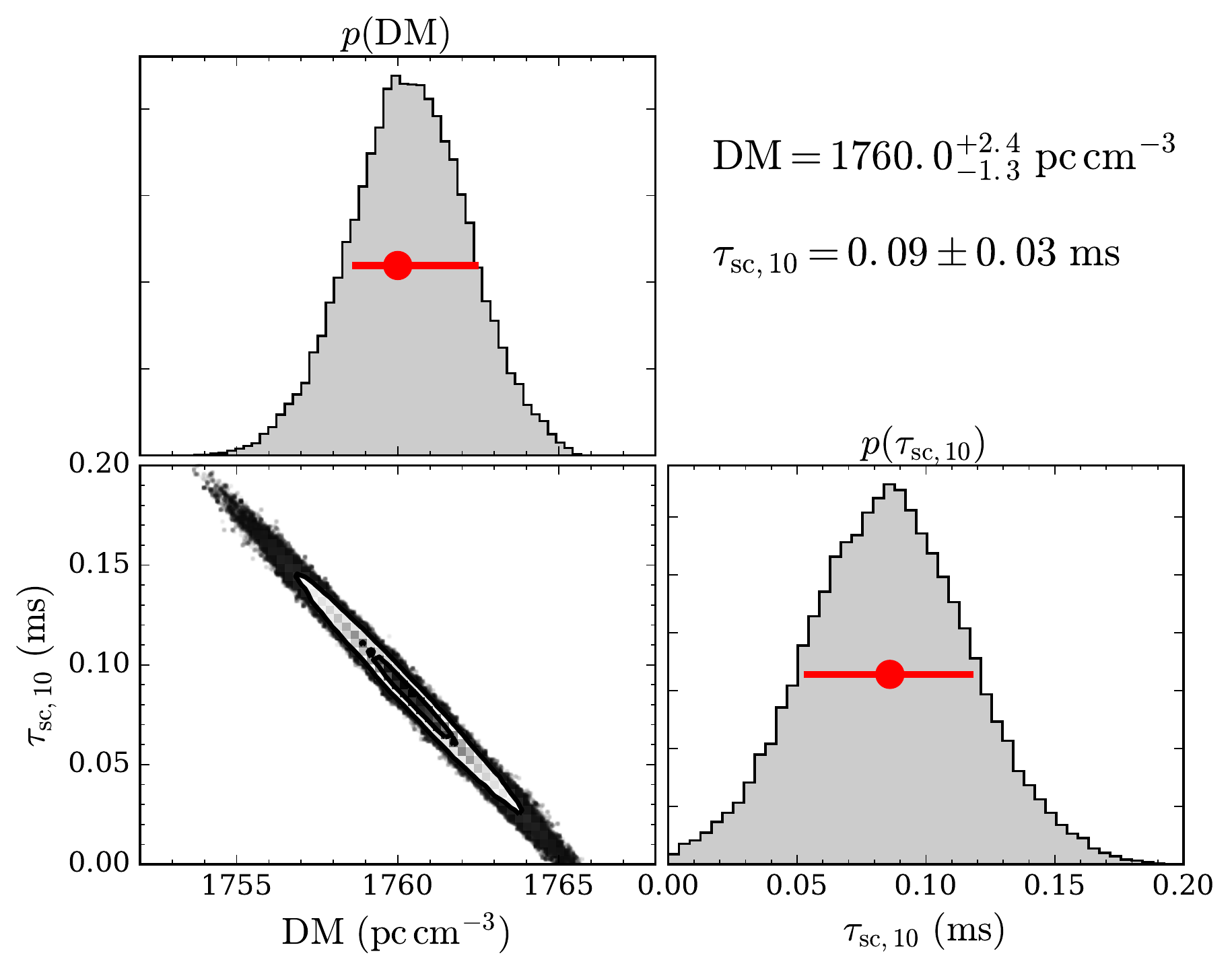}
\caption{Marginalized posterior distributions for the global fit 
         using 38 single pulses.  The two dimensional distribution 
         in the lower left gives the joint posterior distribution of 
         $\rm DM$ and $\tau_{\rm sc,10}$ marginalized over all 38 
         offset terms $\{t_{0,i}\}$.  The grey histograms give the 
         fully marginalized posterior distributions for the 
         $\rm DM$ (top left) and $\tau_{\rm sc, 10}$ (bottom right).
         The peak value and compact innermost 68\% of the posteriors 
         for each parameter are indicated by the red marker and bar.
         }
\label{fig:corner}
\end{figure}

The global fit for all 38 pulses gives 
$\rm DM = 1760.0^{+2.4}_{-1.3}~{\rm pc~cm}^{-3}$ and 
$\tau_{\rm sc, 10} = 0.09 \pm 0.03 ~\rm ms$.  
Figure~\ref{fig:corner} shows the joint posterior 
distribution $p(\rm DM, \tau_{\rm sc, 10})$ marginalized 
over all $\{t_{0,i}\}$ and the fully marginalized posteriors 
$p({\rm DM})$ and $p(\tau_{\rm sc,10})$.  The best fit values 
and uncertainties for $\rm DM$ and $\tau_{\rm sc,10}$ are taken 
as the maximum and most compact inner 68\% of the fully marginalized 
posteriors for each parameter. 

\begin{figure}[hp]
\centering 
  \includegraphics[width=\textwidth]{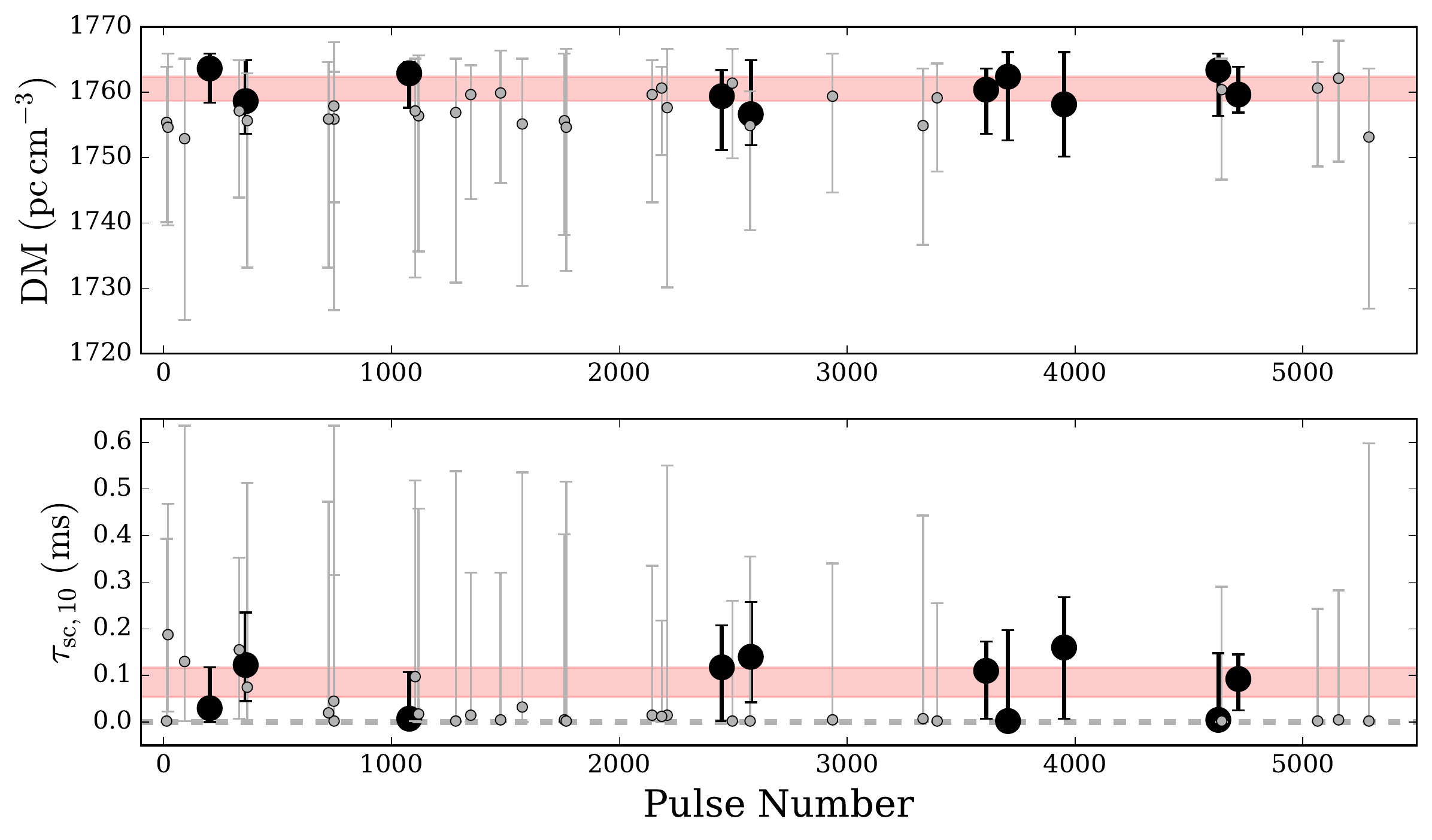}
\caption{Individual fits for the dispersion measure (top) and 
         10~GHz scattering time (bottom) for each of the 38 
         pulses used in the global fit.  The points give the 
         maximum of the fully marginalized posterior for each 
         parameter and the errorbars denote the innermost compact 
         68\% region. For clarity, the parameter values for pulses 
         with the most constraining individual fits 
         ($\rm \delta DM < 10~{\rm pc~cm}^{-3}$) are shown with large 
         black markers.  Parameters from other pulses are shown as 
         smaller grey markers.
         The red bands give the global fit measurements of 
         $\rm DM = 1760.0^{+2.4}_{-1.3}~{\rm pc~cm}^{-3}$ and 
         $\rm \tau_{\rm sc, 10} = 0.09 \pm 0.03 ~ms$.  
        }        
\label{fig:param_fits}
\end{figure}

Figure~\ref{fig:param_fits} shows the individual fits of 
$\rm DM$ and $\tau_{\rm sc,10}$ for each of the 38 pulses.  
The best fit values and uncertainties of 
$\rm DM$ and $\tau_{\rm sc,10}$ for each pulse are taken as the 
maximum and most compact inner 68\% of the fully marginalized 
posteriors for each parameter.  All of the individual pulse fits are 
consistent with constant values for  $\rm DM$ and $\tau_{\rm sc,10}$ 
over the course of the 6~hour observation.

\subsection{Comparison with Previous Results}
\label{ssec:compare}
Shortly after the discovery of radio pulsations from \gcmag, several 
measurements of the dispersion and scattering were made.  
\citet{efk13} measured the dispersion measure of \gcmag\ to be 
$\rm DM = 1778\pm 3~{\rm pc~cm}^{-3}$ from pulsar timing observations 
over a frequency range of 2.5--8.5~GHz.  \citet{sle14} conducted a 
multifrequency study of \gcmag\ using multiple telescopes to measure 
the parameters of the scattering law.  They found a 1~GHz scattering 
time of $\tau_{\rm sc,1} = 1300 \pm 200~\rm ms$ and a scattering index 
of $\alpha_{\rm sc} = -3.8 \pm 0.2$. \citet{ppe15} observed \gcmag\ with 
the GBT in two observing bands to cover 1.4--2.4~GHz and used a 
wide-band model to simultaneously fit the scattering and dispersion 
parameters. Over 28~days of observing, they measured values ranging 
from $\rm DM \approx 1770 - 1800~{\rm pc~cm}^{-3}$ and showing an 
apparent variability in both time and frequency. More recently, 
\citet{dep18} presented the results of the long-term 
monitoring campaign of \gcmag\ that began with the \citet{efk13} 
observations.  Observing over the frequency range of 2.5--8.5~GHz 
over a four year span, they found that the dispersion measure was 
consistent (at the $2\sigma$ level) with a constant value of 
$\rm DM = 1762 \pm 11~{\rm pc~cm}^{-3}$.
Finally, \citet{pmp18} reported a large 
($\tau_{\rm sc, 8.4} \approx 6~\rm ms$) and variable scattering time 
at 8.4~GHz with the Deep Space Network 70-m telescope DSS-43.

Our measurement of a 10~GHz pulse broadening time of 
$\tau_{\rm sc,10} = 0.09 \pm 0.03 ~\rm ms$ is consistent with the 
$\tau_{\rm sc,10} = 0.1-0.3~\rm ms$ expected at 10~GHz from the 
\citet{sle14} scattering relation, but is much less than the 
$\tau_{\rm sc, 10} \approx 3~\rm ms$ expected from the \citet{pmp18} 
result.  Our DM measurement of 
$\rm DM = 1760.0^{+2.4}_{-1.3}~{\rm pc~cm}^{-3}$ is consistent 
with the $\rm DM = 1762 \pm 11~{\rm pc~cm}^{-3}$ value seen by 
\citet{dep18} over a four year span.  However, both our value and 
that of \citet{dep18} are $\Delta \rm DM \approx 10-40~{\rm pc~cm}^{-3}$ 
smaller than the early measurements by \citet{efk13} and \citet{ppe15}.  
The discrepancies in DM and scattering time are discussed in 
Section~\ref{ssec:disc_dm_tsc}.

\section{Discussion}
\label{sec:discussion}
We have conducted a detailed study of single pulses from the 
radio-emitting magnetar \gcmag\ using the VLA in its phased-array 
pulsar mode at 7--12~GHz.

\subsection{Profile Evolution}
\label{ssec:disc_prof_evol}
We have studied both the time and frequency evolution of 
the average profile of \gcmag.  Using two 6.5~hour 
observations on consecutive days, we found that the average pulse 
profile was stable on $\sim$day timescales.  Comparison with 
additional phased VLA observations at 8.7~GHz from July~2013 to 
February~2015, shows that the average pulse profile of \gcmag\ 
changes on longer timescales.  This profile variability is 
consistent with previous observations of \gcmag\ over a range 
of frequencies \citep{lak15, ysw15, ywm18, tek15, tde17} and 
with studies of other radio-emitting magnetars \citep{ccr07}, 
which suggests a magnetospheric origin.  

Using 5~GHz of simultaneous bandwidth from a single epoch, 
we also found that the profile is fairly stable over a frequency 
range of 7--12~GHz ($\Delta \nu / \nu \approx 0.5$), showing only 
a slight narrowing of components with increasing frequency.  This 
modest evolution is consistent with what is seen in radio pulsars 
at comparable frequencies \citep{kxj97, jkm08}.  Radio-emitting 
magnetars (including \gcmag) have shown large profile changes 
(e.g., the appearance and disappearance of components) over 
frequency ranges of $\sim\! \rm GHz$ \citep{ksj07, tek15}. 
However, \gcmag\ has also been observed with a stable pulse 
profile from $2-8~\rm GHz$ \citep{tde17}, so the frequency 
evolution may also be time-dependent.

\subsection{Single Pulses}
\label{ssec:disc_sp}
The wide ($\approx 700$~ms) profile of \gcmag\ in our observations 
is comprised of much narrower ($\sim 1-10$~ms) single pulses.  
This spiky pulse emission is uncommon in radio pulsars, but 
appears to be characteristic of radio-emitting magnetars 
\citep{ksj07,lbb12}.  To study these pulses, we used a 
matched-filter technique to characterize the pulses in 
each profile component for every available rotation of 
\gcmag\ in our data.  Comparing the occurrence, amplitude, and 
arrival time of pulses in each profile component, we find no 
correlation in the amplitude or phase of pulses occurring in 
different profile components.  However, we do find a statistically 
significant over-abundance of pulses occurring during the same rotation 
in both components C1 and C2, possibly suggesting a common or related 
origin for pulses in these two profile components.
We also measured the frequency correlation of the single pulse 
jitter of sub-pulses in each of the four profile components.  
Using data from four 1~GHz sub-bands, we found that the jitter 
is $\approx 100\%$ correlated over the full $7-12$~GHz VLA band.

\subsection{Dispersion Measure and Scattering}
\label{ssec:disc_dm_tsc}
Variations in the dispersion measure and scattering time of \gcmag\ 
probe the inhomogeneities of the distribution of free electrons along 
the line of sight to the Galactic center.  Measuring the magnitude and 
timescale of these variations can help disentangle the 
contributions to the DM and scattering within the Galactic center 
from those occurring along the line of sight in the Galactic plane.  
Any significant variation in the scattering time would also have 
important implications for strategies to find pulsars near \sgra.  
Our single epoch (MJD~56915) measurement of the DM and scattering 
time cannot say much about variability itself, but is useful in the 
context of other published measurements.

\subsubsection{Scattering Variations}
\label{sssec:tsc_var}
Our measurement of the 10~GHz scattering time on MJD~56915 
is consistent with the values measured by \citet{sle14} 
from MJD~56418-98, but this does not rule out the possibility 
that the scattering time is variable.  The relatively large 
uncertainty in our measurement is such that it may differ from 
the \citet{sle14} relation by a factor of a few.  Furthermore, 
it could be the case that scattering is sporadically enhanced 
due to small scale features in the Galactic center. Future 
attempts to measure the scattering in single pulses should 
note the intrinsic asymmetry in some of the pulses we have 
observed (Figure~\ref{fig:dspec_plots}).  Had we ignored the 
frequency dependence of the scattering time and just fit a 
Gaussian convolved with an exponential, we would have 
mistaken this intrinsic structure for scattering times as 
high as several milliseconds.

\subsubsection{DM Variations}
\label{sssec:dm_var}
\citet{dep18} presented the DM of \gcmag\ for over 1500~days 
(starting soon after the detection of radio pulsations), with 
measurements at 2.5, 4.85, and 4-8~GHz. Our measurement of 
$\rm DM = 1760.0^{+2.4}_{-1.3}~{\rm pc~cm}^{-3}$ on MJD~56915 is 
consistent with the most precise single epoch measurement reported 
by \citet{dep18} of  $\rm DM = 1765 \pm 4~{\rm pc~cm}^{-3}$ 898 days 
later (MJD~57813). All of the high frequency 
(4.85, 4-8~GHz) DM measurements from \citet{dep18} also appear 
consistent with a constant value of 
${\rm DM} \approx 1760~{\rm pc~cm}^{-3}$, 
although most of the measurements have uncertainties of 
$\sigma_{\rm DM} \approx 15-25~{\rm pc~cm}^{-3}$ so variations 
below this level cannot be excluded.

Our DM measurement is $\Delta \rm DM \approx 10-40~{\rm pc~cm}^{-3}$ 
smaller than lower frequency ($\nu \approx 1-2~\rm GHz$) measurements 
in the first $\approx \!\! 400$~days afer radio pulsations were detected 
from \gcmag. 
\citet{efk13} used pulsar timing observations at 2.5 and 8.5~GHz and 
found $\rm DM = 1778\pm 3~{\rm pc~cm}^{-3}$ shortly after the first 
detection of radio pulsations (MJD~56414).  \citet{ppe15} found a 
range of $\rm DM \approx 1770 - 1800~{\rm pc~cm}^{-3}$ using the 
GBT in two observing bands to cover 1.4--2.4~GHz over a 28 day 
span from MJD~56488-516.  The 2.5~GHz DM measurements from 
\citet{dep18} over the first $\approx \!\! 400$~days after the magnetar 
radio detection also appear to be systematically higher, though the 
uncertainties are large. 

There are a few possible explanations for the discrepancy between 
our DM and the early low-frequency measurements.  One possibility 
is that the DM decreased by $\Delta \rm DM \approx 10-40~{\rm pc~cm}^{-3}$ 
over the $\approx \!\! 400$ days between the radio detection and our 
measurements.  Such a change could occur if the magnetar travels far 
enough through the dense ionized gas found in the Galactic center.  
Taking a line of sight velocity to be comparable to the transverse 
velocity of $v_\perp = 236 \pm 11 ~{\rm km~s}^{-1}$ measured by 
\citet{bdd15}, \gcmag\ travels a distance of 
\begin{equation}
  \Delta L \approx 2 \times 10^{-4}~{\rm pc}~
    \left(\frac{v_{\rm los}}{200~{\rm km~s}^{-1}}\right) 
    \left(\frac{t}{400~{\rm days}}\right)
\end{equation}
in the 400~days between $\rm DM$ measurements.  In order to 
fully explain the difference in $\rm DM$, the mean electron 
density needs to be 
$\bar{n}_{\rm e} = \Delta {\rm DM} / \Delta L \approx 10^5~{\rm cm}^{-3}$. 
This value is just within the range of electron densities 
($n_{\rm e} \approx 0.2-1 \times 10^5~{\rm cm}^{-3}$) estimated 
along Sgr~A~West towards \gcmag\ by \citet{zbm10} based on radio 
recombination line measurements.  However, Sgr~A~West has an estimated 
depth of $d_{\rm W} \sim 0.1~\rm pc$ \citep{ferriere12}.  For the 
contribution to the observed $\rm DM$ to be 
$\rm DM_w \lesssim 100~{\rm pc~cm}^{-3}$ (for consistency with 
other Galactic center pulsars), \gcmag\ could only be 
$d \lesssim 0.001~\rm pc$ within Sgr~A~West.  
Outside Sgr~A~West, the typical electron density of the warm ionized 
gas in the central cavity is 
$n_{\rm e, c} \sim 10^3~{\rm cm}^{-3}$ \citep{ferriere12}, which  
would only produce $\Delta \rm DM \approx 0.2~{\rm pc~cm}^{-3}$. 
For the observed $\Delta \rm DM$ to be real, then, \gcmag\ would 
need to be located just barely within the densest parts of 
Sgr~A~West.  While not impossible, this seems unlikely.

Another possibility is that the measured DM depends on the 
observing frequency. \citet{css16} describe how frequency 
dependent DMs can arise from multipath propagation in the 
turbulent ISM.  Basically, the measured DM at a given 
observing frequency is the average of many paths that pass 
through the scattering disk.  Since the scattering disk is 
frequency-dependent ($\theta_{\rm sc} \propto \nu^{-2}$), 
different observing frequencies will sample different 
paths through the ISM. However, the predicted offsets between 
DMs measured at 2~GHz and 4, 6, or 10~GHz are only 
$\delta {\rm DM} \sim 1~{\rm pc~cm}^{-3}$, so this effect 
is likely insufficient to make up the difference.

The final possibility is that the difference is the result of 
systematic biases in the different methods for measuring the 
DM and scattering.  We have used a collection of bright and 
narrow single pulses to make our measurements, but the earlier 
lower frequency measurements all used integrated profiles. 
As shown in Figure~\ref{fig:multi_epoch}, the average profiles 
can have widths of $W_{\rm avg} \gtrsim 100~{\rm ms}$ and 
sometimes show multiple (possibly overlapping) profile components. 
Depending on the method used, this could potentially affect the 
DM and scattering measurements.  For example, fitting a single 
Gaussian convolved with an exponential scattering tail to an 
overlapping double peaked profile could result in a measured 
DM that is incorrect.  By instead measuring the time delays 
of narrow single pulses, we should have avoided these issues. 
By jointly fitting many pulses, we further reduce the effect 
of individual pulse shapes.  While single pulse fitting may 
have its own biases, they are likely different than those 
encountered in the average profile fitting.  We consider this 
the most likely explanation for the difference in observed DMs, 
but a long-term campaign to measure both DM and scattering at 
lower frequencies is likely needed to resolve this issue.

\acknowledgments
We thank V. Dhawan, K. Sowinski, and M. Rupen for their assistance in 
the early stages of the phasing effort for the VLA.
The National Radio Astronomy Observatory is a facility of the National 
Science Foundation operated under cooperative agreement by Associated 
Universities, Inc.
SC, JMC, PD, TJWL, and SMR are members of the NANOGrav Physics 
Frontiers Center, which is supported by the National Science Foundation 
award number 1430284.
RSW acknowledges financial support by the European Research Council (ERC) 
for the ERC Synergy Grant BlackHoleCam under contract no. 610058.
Part of this research was carried out at the Jet Propulsion Laboratory,
California Institute of Technology, under a contract with the National
Aeronautics and Space Administration.

\facility{VLA}

\software{PRESTO \citep{ransom2001}, DSPSR \citep{vsb11}, 
          PyPulse \citep{lam17}, 
          emcee \citep{fhl13}, 
          numpy \citep{numpy}, matplotlib \citep{matplotlib}}

\bibliography{gcmag}
\end{document}